\documentclass[tighten,times]{aastex631}
\usepackage[T1]{fontenc}
\usepackage{amsmath}
\usepackage{hyperref}
\usepackage{appendix}

\usepackage{units}
\usepackage{bm}

\newcommand{\response}[1]{{#1}}

\received{AAS46016}

\submitjournal{ApJ 2nd revision, under review}

\newcommand{\BVF}{Brunt-V\"ais\"al\"a-frequency\ }
\newcommand{\BVFs}{Brunt-V\"ais\"al\"a-frequencies\ }

\newcommand{\SH}{Solberg-Høiland\ }

\shortauthors{Klahr}
\shorttitle{Thermal baroclinic instabilities in accretion disks: Linear Theory}

\begin{document}



\title{Thermal baroclinic instabilities in accretion disks I:\\
Combined dispersion relation for Goldreich-Schubert-Fricke Instability and Convective Overstability in disks around young stars}



\correspondingauthor{Hubert Klahr}\email{klahr@mpia.de}
\author[0000-0002-8227-5467]{Hubert Klahr}
\affil{Max Planck Institut f\"ur Astronomie, K\"onigstuhl 17, 69117, Heidelberg, Germany}




\begin{abstract}
This paper discusses the Goldreich-Schubert-Fricke instability (GSF) and the convective overstability (COS) in the context of baroclinic thermal instabilities in rotating disks around young stars. The vertical shear instability (VSI) is a global extension of the GSF that affects geometrically thin disks but follows the same stability criterion. The COS, on the other hand, also possesses a twin for stellar interiors, specifically, Shibahashi's vibrational stability of rotating stars. We derive a combined dispersion relation for GSF and COS with arbitrary cooling times for local perturbations and determine a new stability criterion beyond the \SH criterion. \response{The paper shows that in extension to the stability criterion for the vertically unstratified case ($N^2_R > 0$), one also needs a barotropic disk structure to ensure stability towards COS modes.} We demonstrate that a baroclinic disk atmosphere always has a buoyantly unstable direction, although not necessarily in the radial nor vertical direction. The paper predicts that for cooling times longer than the critical cooling time for VSI, GSF modes will always be accompanied by COS modes of similar growth rate. The numerical companion paper II tests the predictions of growth rates from this paper.
\end{abstract}


\noindent
\section{Introduction}
The disk around a young star may be subject to a range of magnetic and non-magnetic effects driving angular momentum transfer and generating turbulence \citep{Lesur2023}, both of which have a major impact on the prospects of planet formation \citep{Klahr2018}.
From current observations, it is not clear if there is a dominating effect shaping disk structure and if this is a magnetic or pure hydrodynamic effect, or whether it is the same one for all locations in the disk \citep{Lesur2023}. It has been shown that for sufficient ionization and coupling of gas with the magnetic fields, the magneto-rotational instability (MRI) \citep{Balbus1991, 1991ApJ...376..223H} with its fast growth rates will dwarf hydrodynamic instabilities \citep{Lyra2011}. However, the MRI can easily be suppressed by non-ideal magneto-hydrodynamical (MHD) effects related to the low gas temperature and presence of dust and ice grains in the planet-forming regions of disks around young stars. Therefore, both non-ideal MHD as well as pure hydrodynamic instabilities deserve similar attention in our goal to better understand the structure of observed disks as well as the conditions for planet formation. In a disk, several independent instabilities may be operating at different heights and radial locations, possibly even at the same location, only separated by length and time scales. For instance, large scales may be subject to magnetic winds, whereas small scales decouple from the magnetic fields and could experience some action of hydrodynamical instabilities.
The disk around a young star may be subject to a range of magnetic and non-magnetic effects driving angular momentum transfer and generating turbulence \citep{Lesur2023}, both of which have a major impact on the prospects of planet formation \citep{Klahr2018}. 


Several purely hydrodynamical instabilities are currently discussed in the context of dusty gas disks around young stars \citep{Lesur2023}. A pair of instabilities are nonlinear and inherently non-axisymmetric: The Zombie Vortex Instability (ZVI) \citep{Marcus2015,Marcus2016} and the stratorotational instability (SRI) \citep{Shalybkov2005,LeBars2007}. Both properties, non-linearity and non-axisymmetry, make them rather challenging both for analytic treatment as well as for numerical simulations. Also, the Rossby wave instability should be mentioned, which is a linear yet radially global instability \response{relying on a local extremum in vorticity} \citep{Lovelace1999}.

But there are also 
two linear instabilities \response{for axisymmetric modes operating in disks with a radially monotonous pressure and vorticity profile}: Convective Overstability\footnote{We follow the convention of \citet{Tassoul2000}: The plural "instabilities" can refer to non-oscillatory instabilities as well as to overstabilities, which show a growth of amplitude for oscillatory modes.} (COS) \citep{Klahr2014,Lyra2014} and Vertical Shear Instability (VSI) \citep{Urpin1998,Urpin2003,Nelson2013,Barker2015}. Both possess linear growing modes, which makes a detailed stability analysis feasible. \response{The COS has never been properly studied in disks with vertical stratification, which is the primary goal of the present paper.}




\begin{deluxetable}{lll}[tb!]
\tabletypesize{\footnotesize}
\tablecaption{Used symbols and definitions:\label{tab:usedSym}}
\tablehead{\colhead{Symbol} & \colhead{Definition }& \colhead{Description}}
\startdata
$R$,$z$,$\varphi$         &                        & cylindrical coordinates   \\
$\rho, P$      &            & density, pressure\\
$c_v$      &                          & specific heat \\
$E, T$      &  $E = c_v \rho T$          & internal energy, temperature\\
$\gamma$      &                          & adiabatic index \\
$S$      &      $=P \rho^{-\gamma}$      & specific entropy proxy\\
$\mathrm{d} s$      &      $=c_v \mathrm{d} \log S$      & specific entropy\\
$p$         &     $ = \frac{ {\rm d\log}\,\rho(R,0)}{{\rm d\log} R}  $  & global density gradient \\
$q$         &    $ = \frac{{\rm d \log} \, T(R,0)}{{\rm d \log} R}     $ & global temperature gradient\\
${\bf a}$     &    $ =\nabla {\rm \log} \rho   $ & local density stratification\\
${\bf b}$     &    $ =\nabla {\rm \log} P   $ & local pressure stratification\\
${\bf s}$     &    $ =\nabla {\rm \log} S   $ & local entropy stratification\\
$\Omega (R,z)$      &            & Rotational frequency\\
$c$      &      $=P / \rho$        & isothermal speed of sound\\
$H$      &          $=c / \Omega$                     & pressure scale height\\
$h$      &          $=H/R$                     & aspect ratio\\
$\tau$      &                                 & radiative cooling / thermal diffusion time\\
$\tau \gamma$      &                          & thermal relaxation time for constant pressure\\
$\tau^*$      &      $=\tau \gamma \Omega$    & dimensionless thermal relaxation time\\
$\tau_\mathrm{c}$      &      $h |q| \frac{1}{\gamma -1} \Omega^{-1}$                         & critical $\tau$ for VSI\\
$\tau_\mathrm{GSF}$      &      $\frac{1}{2} \frac{H}{z}\frac{H}{R} |q| \frac{1}{\gamma -1}\Omega^{-1}$                         & critical $\tau$ for GSF\\
$k_R,k_z$  &                          & radial, vertical wave number\\
$\mathbf{k}$  &                          & wave number vector\\
$\boldsymbol{a}$  &    $\boldsymbol{a}\cdot \mathbf{k} = 0$                     & direction vector for velocity perturbation\\
$\kappa_R^2$      &       $=\frac{1}{R^3}\partial_R \Omega^2 R^4$                    & radial angular momentum gradient: epicyclic frequency \\
$\kappa_z^2$      &         $=\frac{1}{R^3}\partial_z \Omega^2 R^4$                  & vertical angular momentum gradient\\
$\kappa_\mathbf{k}^2$      &    $=\frac{k^2_z}{k^2}\left(\kappa_R^2 - \frac{k_R}{k_z}\kappa_z^2\right)$                       & oscillation frequency (OF) for $\mathbf{k}$\\
$N^2$      &    $= - \frac{1}{\rho \gamma c_v}\nabla P \nabla K$                & buoyancy frequency (BF)\\
$N_R^2,N_z^2$      &       & radial, vertical BF\\
     &                    & aka \BVF\\
$N_\mathbf{k}^2$      &   $=\frac{N_R^2 \left(1 -\frac{b_z k_R}{b_R k_z}\right) k_z^2 + N_z^2 \left(1 - \frac{b_R k_z}{b_z k_R}\right) k_R^2 }{k^2}$                 & buoyancy frequency for $\mathbf{k}$\\
$N_-^2$      &    $=\min(N_\mathbf{k}^2)_\mathbf{k}$                & lowest local BF\\
$N_+^2$      &    $=\max(N_\mathbf{k}^2)_\mathbf{k}$                & largest local BF\\
$\kappa_-^2$      &    $=\min(\kappa_\mathbf{k}^2)_\mathbf{k}$                & lowest local OF\\
$\kappa_+^2$      &    $=\max(\kappa_\mathbf{k}^2)_\mathbf{k}$                & largest local OF\\
%
$\Gamma_\mathrm{VSI}$      &    $=\frac{H}{R}|q| \Omega$                & typical VSI growth rate\\
$\Gamma_\mathrm{GSF}(\tau < \tau_c)$      &    $=\frac{|z|}{R} \frac{|q|}{2}\Omega$                & GSF growth rate for $\tau \rightarrow 0$\\
$\Gamma_\mathrm{GSF}$      &    $=\frac{q^2}{4}\frac{\gamma }{\gamma-1}\,\,\, \frac{1}{\tau^*_\mathrm{GSF}  + \tau^*} \, \Omega$                & approx. GSF growth rate\\
$\Gamma_\mathrm{COS}$      &    $=\frac{q^2}{8}\frac{\gamma}{\gamma-1}\,\, \frac{\tau^*}{1 + \tau^{*2}} \,\Omega$                & approx COS growth rate\\
\enddata
\end{deluxetable}

The subcritical baroclinic instability (SBI), in which a global radial entropy gradient generates quasi-two-dimensional anti-cyclonic vortices parallel to the midplane \citep{Klahr2003,Klahr2004} that are further amplified at the proper thermal relaxation rate \citep{Petersen2007a,Petersen2007b,Lesur2009,Lyra2011,Raettig2013}, depends similarly on the radial entropy stratification and proper thermal relaxation rate as the COS in vertically unstratified disks. But a linear stability analysis of SBI is not feasible due to its non-linear nature and radial shear complicating any simple analysis by introducing time-dependent wave vectors \citep{Klahr2004}. Both SBI and COS have been studied exclusively in non-stratified disk models, where their occurrence relies on the same conditions. Thus, one might assume that SBI is the nonlinear stage of the COS. However, before reaching such a conclusion, both instabilities need to be investigated in more realistic configurations, including vertical stratification.

The fact that the COS has been studied solely in vertically unstratified disk atmospheres so far has the benefit of suppressing the VSI, which requires some form of vertical stratification in density and pressure. A vertically unstratified disk exhibits a barotropic background state, wherein the pressure and density gradients are parallel and pointing radially. Hence, there is a unique radial Brunt-Väisälä frequency ($N_R^2$), which is either positive, indicating stable stratification, or negative, implying super-adiabatic or potentially convective unstable stratification. The vertical Brunt-Väisälä frequency ($N_z^2$) is always zero without stratification. In vertically stratified disks, except at the midplane, the gradients of density $\mathbf{\nabla}\rho$ and pressure $\mathbf{\nabla}P$ generally do not align, constituting the baroclinic state of the disks. Baroclinicity in a rotating system results in vertical shear in the rotation profile $\Omega(z)$, also known as the "thermal wind" in Earth's atmosphere \citep{Holton2012-kg}. Only vertically stratified disks that are globally at the same temperature or specific entropy are barotropic and would not possess vertical shear. However, such disk structures are highly improbable. Therefore, the natural state for disks irradiated by the central star is to possess a baroclinic structure:

\begin{equation}
    \mathbf{\nabla}P \times \mathbf{\nabla}\rho \ne 0,
\end{equation}
from which the vertical shear $\partial_z \Omega$ can be determined. As we will see later, it is convenient to directly express the vertical shear as gradient of the square of the specific angular momentum $j = \Omega R^2$:
\begin{equation}
    \kappa_z^2 := \frac{1}{R^3} \partial_z j^2 = R \partial_z \Omega^2 = \frac{1}{\rho^2}\nabla \rho \times  \nabla P.
    \label{Eq:2}
\end{equation}
For vertically isothermal disks with a radial temperature gradient of $q={{\rm d \log} \, T}/{{\rm d \log} R}$, this leads to a vertical shear \citep{Nelson+2016}
\response{notably independent from the local temperature of the disk. In other words only the logarithmic gradient matters, but not the local pressure scale H/R:}
\begin{equation}
 \partial_z v_\varphi = \frac{q}{2} \frac{z}{R} \Omega.
 \label{Eq:verticalshear}
\end{equation}
\response{Thus disks with different $H/R$ but identical $q$ will show the same height dependent shear profile.}


\subsection{Stellar Rotation}
\citet{Tassoul2000} explains in detail how to derive dispersion relations for baroclinic stars with and without thermal relaxation. Without thermal relaxation, one employs the Solberg-Høiland criteria for dynamic stability. For the usual stratification of disks around young stars the Solberg-Høiland criteria are always fulfilled \citep{Ruediger2002}. Yet \citet{Tassoul2000}(Section 3.5, page 82) also discusses {\it thermal instabilities} that arise in dynamically stable baroclinic systems, if they are subject to thermal relaxation, like diffusion of heat via radiative transport.
Under these conditions one finds the stellar structure to be potentially unstable for two different processes, namely a \textit{thermal instability}, first discussed by \citet{Goldreich1967} and \citet{Fricke1968}, as well as a \textit{thermal overstability}: the {"Vibrational Stability of Rotating Stars"} found by \citet{Shibahashi1980}.
As discussed by \citet{Shibahashi1980} the overstable modes can be derived from the dispersion relation by \citet{Goldreich1967}(Eq.\ (32)), who already mention: ``The dispersion relation ... contains many
branches other than the one of relevance to our investigation. 
...
We are interested in
rotational instabilities which arise when the angular momentum per unit mass decreases
outward (from the rotation axis).'', which led them to focus on instantaneous cooling.
\citet{Kato1966} described an ``Overstable Convection in a Medium Stratified in Mean Molecular Weight'', which did not consider the effect of rotation. But then \citet{Shibahashi1980} studied the combined effects of rotation, stratification in chemical composition and thermal relaxation, further investigating the full dispersion relation from \citet{Goldreich1967} (their Eq.\ (32)) to discover said "Overstable Convection".
The same overstability is described in \citet{Knobloch1983}, which they named the {\it ABCD instability} ("Axisymmetric BaroClinic Diffusive instability"). \response{Their dispersion relation Eq.\ (15) includes viscosity $\nu$ and a gradient of chemical composition with diffusivity $\kappa_S$. But for the ABCD instability these effects are not necessary and it is shown that the thermal diffusivity can overcome the stabilizing action of the chemical composition gradient.}
They also point out that a similar instability was already studied as "Diffusive destabilisation of the baroclinic circular vortex" in the work by \citet{McIntyre1970} in a very general context of rotating fluids, including viscosity but without gradients in chemical composition.
In the field of geophysical fluid dynamics and stellar rotation, these thermal instabilities 
rarely find attention \citep{Kitchatinov2014}, because they appear often as weaker than non-axisymmetric dynamical instabilities for a given background stratification and thermal and viscous diffusion processes. \response{\citet{Caleo2016} point out that even so the Prandtl number in the radiative zone of stars as the sun may be smaller than unity, it is not small enough to neglect the effect of viscosity. They conclude:"GSF instability is suppressed in the bulk of the radiative zone
of both the Sun and RGs (red giants) at various evolutionary stages."} 

Stars, as well as the ocean and earth atmosphere, which have primarily rigid rotation, can be unstable to dynamical non-axisymmetric instabilities like the baroclinic instability in the fashion of \citet{Eady1949} and \citet{Charney1947}. These dynamical instabilities have typically larger growth rates than thermal baroclinic instabilities \citep{Tassoul2000}. Additionally, magnetic fields in stars were found to be of bigger importance than thermal baroclinic instabilities \citep{Balbus2001, Menou2004}. \response{Nevertheless, the dispersion relation derived in these papers (e.g.\ Eq.\ (41) in \citep{Balbus2001}) contains our dispersion relation as a subset for the case of negligible viscosity and the absence of a magnetic field. Specifically in \citet{Menou2004} the criterion for COS stability can be found in section 3.2.5.: {\it Requirement $det (2) > 0$ for Stability when $\nu \rightarrow 0$}, including the same necessary stability criterion, yet not discussing the possible oscillatory modes.} 

An analysis of these oscillatory modes can be found in \citet{Urpin2003}. He derives growth rates for the thermal overstability and instability based on the local buoyancy frequency and the gradient of specific angular momentum that are identical to our finding in this paper. Yet, in his work convective stability as derived with the \SH criterion was thought to be sufficient to suppress overstable modes, which in accordance to above mentioned $det(2) > 0$ criterion in \citet{Menou2004} is not sufficient.
In the present paper we show that vertical shear and buoyancy are proportional to another in a baroclinic atmosphere and as a result COS modes experience similar growth rates as GSF in the long cooling time regime, that is a cooling time longer than the critical cooling for VSI \citep{Lin2015}.

\subsection{A note on nomenclature: VSI vs. Kelvin-Helmholtz Instability}
The current name for GSF in the context of disks is VSI, vertical shear instability, which is a bit unfortunate as in the stellar rotation and geophysical communities vertical shear instabilities rely on a different mechanism
\citep{Garaud2021}. What GSF describes would be classified as a centrifugal instability, because it is the \response{radial projection of the angular momentum gradient vector} (see Eq.\ \eqref{Eq.Tassoul:B}) that controls this instability. \response{For that reason, neither the GSF (nor the VSI) has any strictly vertical unstable modes with $k_R = 0$.}

On the other hand, the vertical shear instability in the context of stars consists of non-axisymmetric modes and operates in the vertical direction $k_R = 0$. This shear instability does not rely on the rotation of the system. It is a general shear instability in stratified fluids, aka Kelvin-Helmholtz instability, controlled by the Richardson number $J$ of the flow, which states that the destabilizing shear $\partial_z v_\varphi$ has to overcome the stable stratification expressed as \BVF $N_z^2 > 0$
\begin{equation}
    J  = \frac{N_z^2}{(\partial_z v_\varphi)^2}, 
\end{equation}
\response{where $J > 1/4$ is a sufficient condition for stability \citep{Miles1961,Howard1961}.} Without thermal relaxation the critical Richardson number is reached for $\partial_z v_\varphi^2 > 4 N_z^2$. With vertical shear in a vertically isothermal disk (Eq.\ \eqref{Eq:verticalshear}) and the proper expression for $N_z^2$ \citep{Lin2015}
\begin{equation}
    N^2_z  = -\frac{1}{\gamma} \partial_z P \partial_z S = -\frac{\gamma - 1 }{\gamma} \frac{z^2}{H^2} \Omega^2,
\end{equation}
we find a Richardson number of: 
\begin{equation}
    J  = \frac{4}{q^2} \frac{R^2}{H^2} \frac{\gamma - 1 }{\gamma},
\end{equation}
independent of height. The condition $J<1/4$ puts a constraint on $q$, which is 
\begin{equation}
|q| > 4 \frac{R}{H} \sqrt{\frac{\gamma - 1 }{\gamma}}.
\end{equation}
Even for a warm disk of $H/R \approx 0.1$ this would need an unrealistically strong temperature stratification of $|q| \approx 20 - 30$ depending on $\gamma$.

\response{However, this stability criterion can be compromised by thermal relaxation through radiation or heat conduction, in conjunction with viscosity \citep{Garaud2021,Parker2021}.} In the case of instantaneous cooling, $N^2_z$ ceases to be a constraint, allowing for KHI modes to grow on the order of the vertical shear rate multiplied by the azimuthal wave number $m$. But simultaneously, the radial shear tries to remove any non-axisymmetric structure at a rate proportionally to $m$ because the radial wavenumber changes in first order as $k_r \sim - m \Omega t$, which describes the winding up of spirals \citep{Klahr2004}. As shown by \citet{1965MNRAS.130..125G}, non-axisymmetric perturbations cannot have a simple waveform
because of the effect of
the shearing background on the wave crests. This means for the KHI to develop, the vertical shear has to be stronger than the radial shear, which leads to the condition:
\begin{equation}
|\partial_z v_\varphi| = \frac{|q|}{2} \frac{|z|}{R} \Omega > \frac{3}{2} \Omega.
\end{equation}
As a consequence, the development of the KHI, even for instantaneous cooling, necessitates an exceptionally strong radial temperature gradient:
\begin{equation}
|q| > 3 \frac{R}{|z|},
\end{equation}
which is typically not realized in an accretion disk. A more detailed WKB analysis on radial shear, which suppresses non-axisymmetric instabilities growing over time scales longer than the Keplerian period, can be found in \citet{Knobloch1986}. Additionally, refer to \citet{Shalybkov2005} for a discussion on the stratorotational instability occurring in Keplerian disks with stable vertical stratification, which also results in non-axisymmetric patterns.


To summarize, GSF is an inertial or centrifugal (thermal instability) that remains unaffected by radial shear due to its axisymmetric modes. In contrast, the KHI, as a classical vertical shear instability, does not require rotation and develops non-axisymmetric modes aligned with the flow direction. Although the atmosphere may rotate, radial shear is typically negligible, as seen in scenarios such as a star or planetary atmosphere.

\subsection{Star vs.\ disk}
The situation in a disk is similar to a star in terms of baroclinicity and thus vertical shear, yet one thing is fundamentally different. Whereas stars and planetary atmospheres rotate to first order as solid bodies, disks are subject to Keplerian shear. The shear introduces an important time scale that is the time over which non-axisymmetric perturbations get sheared out and this shear time is on the order of the orbital period. Thus non-axisymmetric instabilities with growth times longer than the orbital period are strongly suppressed. Only instabilities with very strong growth rates, for instance Rossby wave instability (RWI), gravitational instability (GI) and some axial MRI modes can be manifested in Keplerian disks, provided that the right conditions are met. For the RWI one needs a strong local perturbation in the rotation profile, for GI, a very large disk mass at low temperatures, and for MRI, sufficient coupling of magnetic fields to the ionized gas. Note that the analysis of the classical MRI as discussed by \citet{Balbus1991} also relies on axisymmetric modes.

So in disks, with non-axisymmetric hydrodynamic instabilities like the "pure" vertical shear instability and dynamic baroclinic instability being suppressed by radial shear \citep{Knobloch1986}, it opens an avenue for the axisymmetric thermal instabilities, which are according to \citet{Tassoul2000} also baroclinic in nature, yet in contrast to dynamical instabilities dependent on thermal diffusion.
%
\subsection{Outline}
The goal of this paper is to derive the stability criterion for thermal instabilities and apply it to the conditions found in disks. The original plan was to study exclusively Convective Overstability (COS) including vertical stratification, but it quickly became clear that COS and GSF (or respectively VSI) are inseparable for the typical COS thermal relaxation rates on the order of the orbital frequency \citep{Klahr2014}.

For both COS and GSF modes, we derive typical growth rates that can be expected in disks around young stars, such as the solar nebula. These growth rates shall then be tested in nonlinear numerical simulations (see Paper II: \citet{Klahr2023}).

The original VSI work by \citet{Urpin1998,Urpin2003} considers local modes, whereas all recent analytic work on VSI considers vertical global modes \citep{Nelson+2016,Lin2015,Barker2015,Latter2017,Cui2022,Latter2022}. So, to separate the local treatment in the present paper from the global treatment in these papers, I refer to the GSF, by which I mean discussing the local dispersion relation including thermal relaxation, following similar work in the stellar context \citep{Goldreich1967,Fricke1968,Shibahashi1980,Knobloch1983,Balbus2001,Menou2004}. By this definition, GSF is a local instability, whereas VSI is an overstability with radially propagating waves, even though the stability criterion is the same.

In Section 2, we derive the dispersion relation for baroclinic discs using a Boussinesq ansatz and discuss the condition for stability in Section 3. In Section 3, we calculate expected growth rates, both directly via numerically solving the dispersion relation and also by analytic approximations in Section 4. We conclude with a short discussion and outlook to numerical experiments on thermal instabilities in disks in Section 5. A proof that Buoyancy and vertical shear are tightly linked $N^2_- N^2_+ = - \nicefrac{\kappa_z^4}{4}  $ can be found in the appendix.

\section{Stability analysis}
Recent stability analyses for disks often start with the full compressible equations, as for instance by \citet{Ruediger2002} in deriving the Solberg-Høiland criteria, by \citet{Lyra2014} for the study of COS and by \citet{Nelson2013} for the derivation of VSI growth rates. This approach naturally includes sound waves, which can be problematic in the local plane wave ansatz and are usually removed from the dispersion relation \citep{Lyra2014,Nelson2013} to avoid mimicking additional instabilities \citep{MohandasPessah2015}.
Also note that the derivation for the semi-global disk VSI \citep{Nelson2013,Latter2017,Lin2015} one studies the development of modes spanning over the whole vertical atmosphere for a better interpretation of numerical experiments.

But in this semi-global approach one is already neglecting the potentially horizontal COS modes, as the thermal overstable modes in \citet{Tassoul2000} are roughly orthogonal to the GSF modes.
Therefore, in the present work we come back to a local stability analysis in the Boussinesq ansatz of the GSF \citep{Goldreich1967} and VSI \citep{Urpin2003} and leave any discussion on semi-global COS modes for later.

The following derivation can also be found in work on magnetic instabilities in rotating stars \citep{Balbus2001}, with the only difference that we ignore magnetic fields for the moment and assume the absence of viscosity.

One starts with the full compressible set of hydrodynamic equations:
\begin{eqnarray}
&\partial_t& \rho + \mathbf{\nabla} \cdot \left(\rho \mathbf{u}\right) = 0 \nonumber\\
&\partial_t& \mathbf{u} + \left(\mathbf{u} \cdot \mathbf{\nabla}\right) \mathbf{u} = - \frac{1}{\rho}\mathbf{\nabla} P - \mathbf{\nabla} \Phi\nonumber\\
&\partial_t& E + \mathbf{\nabla} \cdot \left(E \mathbf{u}\right) = - P \mathbf{\nabla} \cdot \mathbf{u} - \frac{E_0}{T_0}\frac{T - T_0}{\tau}.
\label{Eq:NonLinear}
\end{eqnarray}
The term for thermal relaxation can also be replaced by an expression for thermal diffusion with diffusivity $\chi$ if we define:
\begin{eqnarray}
- \frac{T - T_0}{\tau} = \chi \mathbf{\nabla}^2 T,
\end{eqnarray}
for suitable boundary conditions. For stellar interiors and geophysical applications the dominant cooling is in fact a diffusive process, which favors the cooling of short wavelength perturbations. In disks around young stars however, the limit for cooling often does not stem from the diffusion of radiation energy, because the mean free path for photons can be larger than the length scale on which the instabilities will develop \citep{Malygin2017}. It is optically thin emission from the dust grains \citep{Lin2015} and the collisions between dust and gas \citep{Malygin2017,Pfeil2019,Muley2023} that define a length-scale-independent cooling time. \response{Recent work on dust size evolution and VSI has found that especially in the outer part of disks, which are accessible by observations, it is the collision timescale that regulates the occurence VSI \citep{Pfeil2023,MelonFuksman2024a,MelonFuksman2024b}.}

The idea for the Boussinesq ansatz (see for instance \citet{Menou2004} for a similar derivation, yet including magnetic fields \response{and viscosity}) is to keep the density evolution due to stratification in entropy and density, while \response{the pressure fluctuations are small}. The new continuity equation is then given as
\begin{eqnarray}
&\partial_t& \log \rho - \frac{1}{\gamma} \mathbf{u} \cdot \mathbf{\nabla} \log S  - \frac{1}{T_0}\frac{T - T_0}{\gamma\tau} = 0,
\label{Eq:buoyancy}
\end{eqnarray}
for which we introduced the specific entropy proxy $S$ here defined via $P_0=S \rho^\gamma$. Additionally we enforce the divergence of the velocities explicitly to zero.
\begin{eqnarray}
 \mathbf{\nabla}\cdot \mathbf{u} = 0.
\end{eqnarray}

The momentum equations in cylindrical coordinates $(R,z,\varphi)$ assuming axisymmetry are
\begin{eqnarray}
&\partial_t& u_R + u_R \partial_R u_R + u_z \partial_z u_R - \frac{u_\varphi^2}{R}= - \frac{1}{\rho}\partial_R P - \partial_R \Phi \nonumber\\
&\partial_t& u_z + u_R \partial_R u_z + u_z \partial_z u_z = - \frac{1}{\rho}\partial_z P - \partial_z \Phi\nonumber\\
&\partial_t& u_\varphi + u_R \partial_R u_\varphi + u_z \partial_z u_\varphi + \frac{u_\varphi u_R}{R} = 0.
\end{eqnarray}
Now we can use a linearization of temperature $T = T_0 + T'$, density $\rho = \rho_0 + \rho'$, pressure $P = P_0 + P'$ and velocity $u_R = u_R'$, $u_z = u_z'$ and $u_\varphi = u_\varphi' + R \Omega$ around an equilibrium solution of a disk or any rotating body under axisymmetry. $\Omega(R,z)$ is the unperturbed rotation profile in which gravity and pressure gradients cancel each other out, thus $u_{R,0} = u_{z,0} = 0$ and all time derivatives are equal to zero. 
Our derivation so far is independent of the actual shape of the rotating body. Only once we plug in specific values for the radial and vertical gradients in density and pressure and include the shape of gravity, then we actually define whether it is a disk, a stellar interior, the earth atmosphere, the ocean, or a lab experiment.

We can express temperature fluctuations in \response{Eq.\ \eqref{Eq:buoyancy}} as density fluctuations because we assume pressure fluctuations to be small, i.e.\ we are not interested in sound waves. Thus from the relation
\begin{equation}
    \frac{P}{P_0} = \frac{T}{T_0}\frac{\rho}{\rho_0} = 0,
\end{equation}
it follows directly
\begin{equation}
    \frac{T'}{T_0} = -\frac{\rho'}{\rho_0},
\end{equation}
and the four linearized equations are: 
\begin{eqnarray}
&\partial_t& \frac{\rho'}{\rho_0} + \frac{1}{\gamma\tau} \frac{\rho'}{\rho_0} -  u_R' \frac{1}{\gamma} \partial_R \ln S_0 - u_z' \frac{1}{\gamma} \partial_z \ln S_0  = 0\nonumber\\
&\partial_t& u_R'  - \frac{\rho'}{\rho_0^2}\partial_R P_0 - \frac{1}{\rho_0}\partial_R P' - 2 u_\varphi' \Omega  = 0\nonumber\\
&\partial_t& u_z'  - \frac{\rho'}{\rho_0^2}\partial_z P_0 - \frac{1}{\rho_0}\partial_z P' = 0\nonumber\\
&\partial_t& u_\varphi' + u_R' \frac{1}{R} \partial_R R^2 \Omega + u_z' R \partial_z \Omega = 0.
\label{eq:Linear}
\end{eqnarray}
Additionally, we can now enforce the velocities to be divergence free 
\begin{eqnarray}
& &\partial_R u_R' + \partial_z u_z'  = 0.
\end{eqnarray}
If we consider a plane wave ansatz for the perturbations $\exp(-i (\omega t - k_R R - k_z z)$,
then the last condition \response{leads to}
\begin{eqnarray}
k_R u_R' + k_z u_z'  = 0.
\end{eqnarray}
\response{We finally have the same set of Equations as in \citet{Menou2004} (Eq. (5),(6),(7),(8) and (11)), if one ignores magnetic fields and sets the viscosity to zero.}
\begin{mathletters}
\begin{equation}
0=\\
\left(
\begin{array}{ccccc}
-i\omega + \frac{1}{\gamma \tau}&  - \frac{1}{\gamma} s_R   &  - \frac{1}{\gamma} s_z & 0 & 0\\
- c^2 b_R  & -i\omega  & 0 &- 2 \Omega & i c^2 k_R\\
- c^2 b_z & 0 &-i\omega  & 0 & i c^2 k_z\\
0 & \frac{\kappa_R^2}{2 \Omega} & \frac{\kappa_z^2}{2 \Omega} &   -i\omega & 0\\
0 & k_R & k_z & 0 & 0 \\
\end{array} \; \; \right) \cdot \left(
\begin{array}{ccc} 
\rho' / \rho_0\\
u_R'\\
u_z'\\
u_\varphi'\\
P'
\end{array} \; \; \right)\;
\label{Eq:Matrix}
\end{equation}
\end{mathletters}
Here we introduced the following \response{definitions} for the local gradients in the entropy and pressure background, which define the \response{baroclinic} structure of the disk (See Appendix A):
\begin{equation}
s_R = \frac{\partial {\rm log} S_0}{\partial R}\,\,\,\, s_z = \frac{\partial {\rm log} S_0}{\partial z} 
\end{equation}
and 
\begin{equation}
b_R = \frac{\partial {\rm log} P_0}{\partial R}\,\,\,\, b_z = \frac{\partial {\rm log} P_0}{\partial z},
\end{equation}
because in this fashion they come in the same units as the wavenumbers $k$, i.e.\ inverse length, and thus are easily comparable. 

The local background state of the disk atmosphere $\rho_0, T_0, S_0, \partial_z \Omega$ is then described from the local values of $a$ and $b$, which can be determined by solving Eqs. \eqref{Eq:NonLinear} for hydrostatic balance in a disk with some global stratification in the midplane density $p = d \log \rho_m / d \log R$ and midplane temperature $q = d \log T_m / d \log R$ \citep{Lin2015}.
The local entropy gradient $\mathbf{s} = \mathbf{b} - \gamma \mathbf{a}$ is defined via the local pressure gradient $\mathbf{b}$ and density gradient $\mathbf{a}$:
\begin{equation}
a_R = \frac{\partial {\rm log} \rho_0}{\partial R}\,\,\,\, a_z = \frac{\partial {\rm log} \rho_0}{\partial z}. 
\end{equation}
We also use the definition of the epicyclic frequency $\kappa_R$, which is proportional to the angular momentum gradient $\partial_R j$:
\begin{equation}
    \kappa_R^2 = \frac{1}{R^3}\partial_R \left(\Omega R^2\right)^2 =  2 \frac{\Omega}{R}\partial_R \Omega R^2.
\end{equation}
For convenience, we define an equivalent expression for the vertical angular momentum gradient $\partial_z j$:
\begin{equation}
    \kappa_z^2 = \frac{1}{R^3}\partial_z \left(\Omega R^2\right)^2 =  2 \Omega \partial_z \Omega.
    \label{Eq:kappaz2}
\end{equation}
This is not to be confused with the vertical oscillatory frequency of an
inclined particle orbit, sometimes called vertical epicyclic frequency.
But in our notation the vector $\bm{\kappa}^2 = \frac{1}{R^3}\mathbf{\nabla} j^2$ is the gradient of the square of specific angular momentum, which will be useful later.
Finally, we use the definition for an isothermal speed of sound $c^2 = P_0 / \rho_0$ to describe the temperature at a given location in the disk atmosphere.

For a rotating star, one finds $\kappa_R^2 = 4 \Omega^2$ and for a Keplerian disk $\kappa_R^2 = \Omega^2$, if one can ignore the effects of radial pressure gradients. The rotation profile in a baroclinic system follows from Eq.\ \eqref{Eq:2} and leads in our notation to
\begin{equation}
    \kappa_z^2 = c^2 \left(a_R b_z - a_z b_R\right) = -\frac{c^2}{\gamma} \left(s_R b_z - s_z b_R\right),
    \label{Eq.kappa_z}
\end{equation}
either in terms of density gradient $\mathbf{a}$ or 
entropy gradient $\mathbf{s}$. 
The determinant in Eq.\ \eqref{Eq:Matrix} has to vanish, which leads to the 3rd order dispersion relation:
\begin{eqnarray}
\omega ^3  + \frac{i}{\gamma  \tau} \omega ^2  + \omega \left[\frac{1}{\gamma k^2} c^2  (k_R b_z -k_z b_R )
   (k_R s_z - k_z s_R ) - \frac{k_z^2}{k^2}  (\kappa_R^2 - \frac{k_R}{k_z}\kappa_z^2)\right] - \frac{i}{\gamma  \tau } \frac{k_z^2}{k^2} (\kappa_R^2 - \frac{k_R}{k_z} \kappa_z^2) = 0,
   \label{Eq:FullDispOmega}
\end{eqnarray}
which can be found in a quasi identical fashion in \citep{Goldreich1967}, which we will discuss later.
One can derive the same dispersion relation from the fully compressible system, yet then it is of 5th order and contains several complex terms that make the further discussion less straight forward \citep{MohandasPessah2015}.

%
%
To obtain a compact expression that also highlights the involved physics, we define a new \BVF for arbitrary direction or respective wavenumber\footnote{In the notation of \citet{Urpin2003} this is the term $\omega_g^2$ \response{Eq.\ (8)}, but we want to stick with the notation of \BVFs.}:
\begin{eqnarray}
 N^2_\mathbf{k}= - \frac{c^2}{\gamma} \frac{(k_R b_z  - k_z b_R) \left(k_R s_z  - k_z s_R  \right)}{k^2} \equiv - \frac{c^2}{\gamma}  \frac{({\bf k} \times {\bf b})\cdot({\bf k}  \times {\bf s})}{k^2}.
  \label{eq:BVTK}
\end{eqnarray}
By writing the terms in this fashion, one recognizes that instead of calculating the product of pressure and entropy gradient, one first makes a projection of both gradients ${\bf b} =\big(\begin{smallmatrix}
  b_R\\
  b_z
\end{smallmatrix}\big)$ and ${\bf s} =\big(\begin{smallmatrix}
  s_R\\
  s_z
\end{smallmatrix}\big)$ individually orthogonal to the direction of the wave vector
${\bf k} =\big(\begin{smallmatrix}
  k_R\\
  k_z
\end{smallmatrix}\big)$ (see also \citet{Tassoul2000}). Thus our $N_\mathbf{k}$ is the potentially complex frequency of internal gravity waves for the given wave vector $\mathbf{k}$ or in other words for perturbations of velocity in the direction perpendicular to $\mathbf{k}$. 
For $k_R = 0$, we find the vertical \BVF $N_z^2 = - (c^2/\gamma) s_z b_z$ and for $k_z = 0$ the radial \BVF $N_R^2 = - (c^2/\gamma) s_R b_R$. We can also introduce these expressions into Eq. (\eqref{eq:BVTK})
\begin{eqnarray}
 k^2 N^2_\mathbf{k} = k_r^2 N_R^2 + k_r k_z \frac{c^2}{\gamma} \left(b_z s_R + b_R s_z \right) + k_z^2 N_z^2,
  \label{eq:BVTK2}
\end{eqnarray}
which shows that $N^2_R$ and $N^2_z$ alone cannot indicate the sign of $N^2_k$.

Likewise, we define an oscillation frequency (OF)\footnote{This is equal to the definition of $Q^2$ \response{in Eq.\ (8)} of \citet{Urpin2003}.} $\kappa^2_\mathbf{k}$ for perturbations in a given direction:
\begin{eqnarray}
\kappa_\mathbf{k}^2 = \frac{k_z^2}{k^2}\left(\kappa_R^2 - \frac{k_R}{k_z} \kappa_z^2\right) = \frac{({\bf k} \times {\bm \kappa^2})({\bf k}  \times {\bf \nabla} R)}{k^2},
\label{eq:2dKAP}
\end{eqnarray}
where we added the last term for an easier comparison with the notation in \citet{Tassoul2000} (see our Eqs.\ (\eqref{Eq.Tassoul:A}) and (\eqref{Eq.Tassoul:B})). \response{Here the gradient of angular momentum in a certain direction perpendicular to $\bf k$ is projected onto the radial part of a displacement ${\bf \nabla} R$ in that same direction perpendicular to ${\bf k}$, which leads to the vanishing of $\kappa_\mathbf{k}^2$ for $k_z = 0$ because of $k_z {\bf \nabla} R = 0$.} 

\response{This notation elucidates that it is not the vertical shear, which becomes unstable, but the radial component of an inclined perturbation, that propagates into a region of lower angular momentum for a larger radius, i.e.\ a centrifugal instability violating Rayleighs criterion.
For radial oscillations with $k_R = 0$ the expression reduces to the radial epicyclic frequency $\kappa_R^2$.}

Thus we find a compact dispersion relation for \response{the radially and vertically stratified case as}
\begin{eqnarray}
\omega ^3  + \omega ^2  \frac{i}{\gamma  \tau} - \omega \left[N_\mathbf{k}^2 + \kappa_\mathbf{k}^2\right] - \frac{i}{\gamma  \tau } \kappa_\mathbf{k}^2 = 0.
\label{Eq:29}
\end{eqnarray}
\response{Removing the vertical stratification, i.e.\ $N^2_z = 0$ and $\kappa^2_z = 0$ leads back to the dispersion relation as derived in \citet{Klahr2014} Eq.\ (21) and \citet{Lyra2014} for radial COS modes:
\begin{eqnarray}
\omega ^3  + \omega ^2  \frac{i}{\gamma  \tau} - \omega \left[N_R^2 + \kappa_R^2\right] - \frac{i}{\gamma  \tau } \kappa_R^2 = 0.
\end{eqnarray}
}

\response{For the further discussion we can rewrite Eq.\ \eqref{Eq:29}} in terms of growth rates instead of frequency $\omega = i\Gamma$ as then all parameters are real:
\begin{eqnarray}
\Gamma^3  + \Gamma ^2  \frac{1}{\gamma  \tau} + \Gamma \left[N_\mathbf{k}^2 + \kappa_\mathbf{k}^2\right] + \frac{1}{\gamma  \tau } \kappa_\mathbf{k}^2 = 0.
\label{q:FullDispOmega}
\end{eqnarray}
\citet{Tassoul2000} similarly 
finds the dispersion relation 
\begin{eqnarray}
\Gamma^3 + \epsilon \Gamma^2 + A \Gamma + \epsilon B = 0,
\label{eq:n3}
\end{eqnarray}
with two complex and one real root related to the growth rates $\Gamma = \pm i \omega_0 + a_\mathrm{T}$ for the oscillatory modes and $\Gamma = b_\mathrm{T}$ for the linear mode.
The thermal relaxation rate here is as $\epsilon = D k^2$ with $D = \chi / \gamma$, the effective thermal diffusivity. The $\epsilon$ prescription is equivalent to our relaxation time $1/(\gamma\tau) = \epsilon$ and can be exchanged arbitrarily. \response{Note that thermal diffusivity $\chi$ and cooling time $\tau$ are defined for a process at constant density. This introduces the adiabatic index $\gamma$ to our equations to allow for expansion and contraction for heating and cooling under constant pressure.}
The factors $A$ and $B$ are given in \citet{Tassoul2000} (section 3.5 page 82) and correspond to our values of $\kappa_\mathbf{k}^2$ and $N_\mathbf{k}^2$ as:
\begin{eqnarray}
A = (\boldsymbol{a} \cdot \mathbf{\nabla}j)(\boldsymbol{a} \cdot \mathbf{\nabla}R) + (\boldsymbol{a} \cdot \mathbf{\nabla}S)(\boldsymbol{a} \cdot \mathbf{\nabla}P) \equiv  \kappa_\mathbf{k}^2 + N_\mathbf{k}^2,
\label{Eq.Tassoul:A}
\end{eqnarray}
and 
\begin{eqnarray}
B = (\boldsymbol{a} \cdot \mathbf{\nabla}j)(\boldsymbol{a} \cdot \mathbf{\nabla}R)  \equiv \kappa_\mathbf{k}^2.
\label{Eq.Tassoul:B}
\end{eqnarray}
\response{\citet{Tassoul2000} does not operate with the wave vector $\bf k$, but with the direction vector of the perturbation $\boldsymbol{a}$, which is perpendicular to $\bf k$ and defined as ${\bf k} \cdot {\boldsymbol{a}} = 0$. We choose a different symbol to prevent confusion with our notation for the density gradient $\bf a$.}

In contrast to \citet{Tassoul2000}, we ignore the effect of stratification of the mean molecular weight $\mu$, i.e.\ chemical gradients, which would have been:
\begin{eqnarray}
\left(\boldsymbol{a} \cdot \frac{\mathbf{\nabla}\mu}{\mu}\right)(\boldsymbol{a} \cdot \mathbf{\nabla}P) =0.
\end{eqnarray}
This is an important consideration for the hydrodynamic stability of stars, potentially suppressing the GSF. It is noteworthy that such a stabilization is also discussed in circumstellar disks for the situation that dust grains have sedimented and are strongly concentrated in the midplane. Thus VSI may be hampered \citep{Lin2019}. Note that the overstable modes, on the other hand, will not be damped by this increase of vertical stratification \citep{Tassoul2000,Shibahashi1980}. A more vigorous discussion of this effect is beyond the scope of this paper.
\begin{figure*}
    \centering
    \gridline{\fig{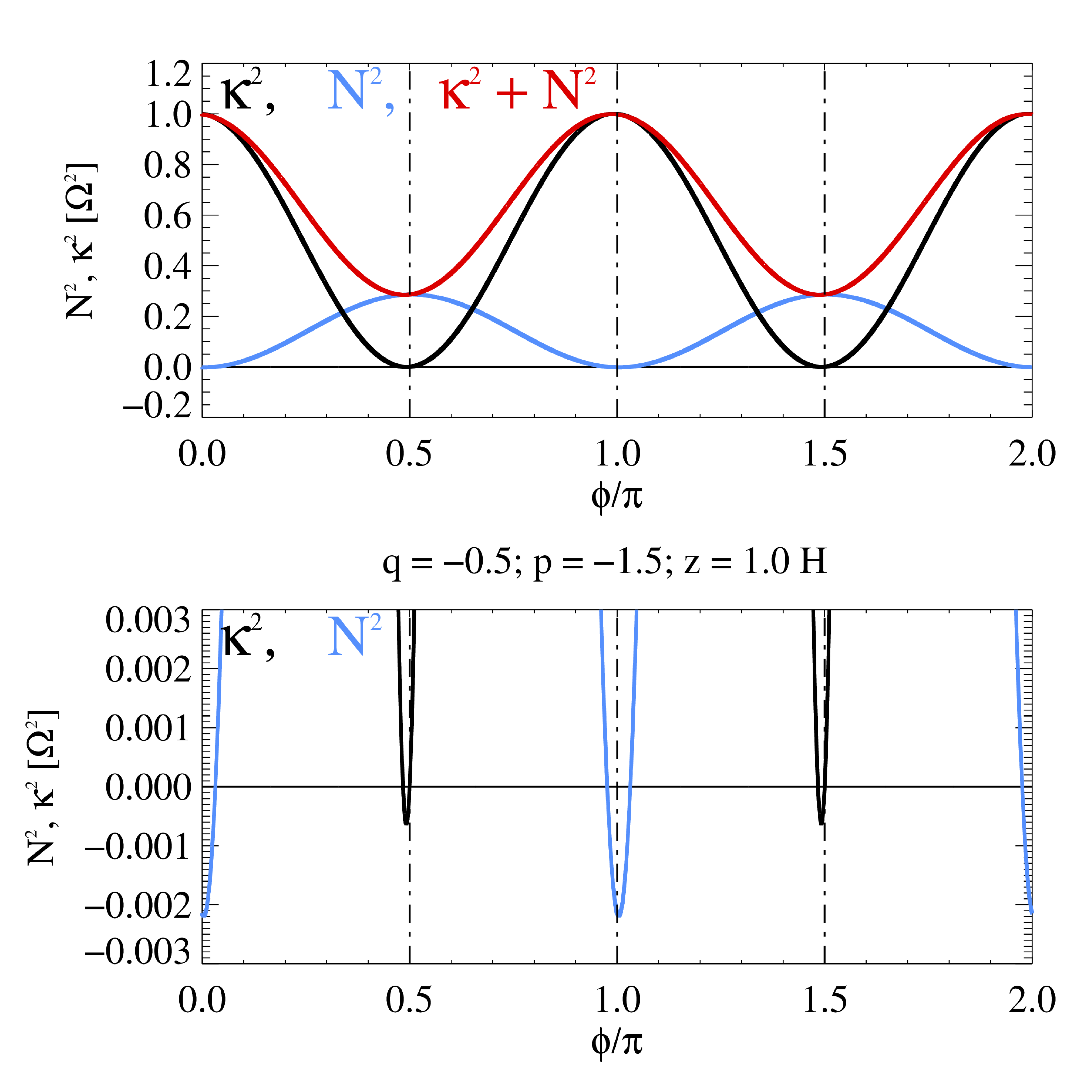}{0.5\textwidth}{(a): h = H}\fig{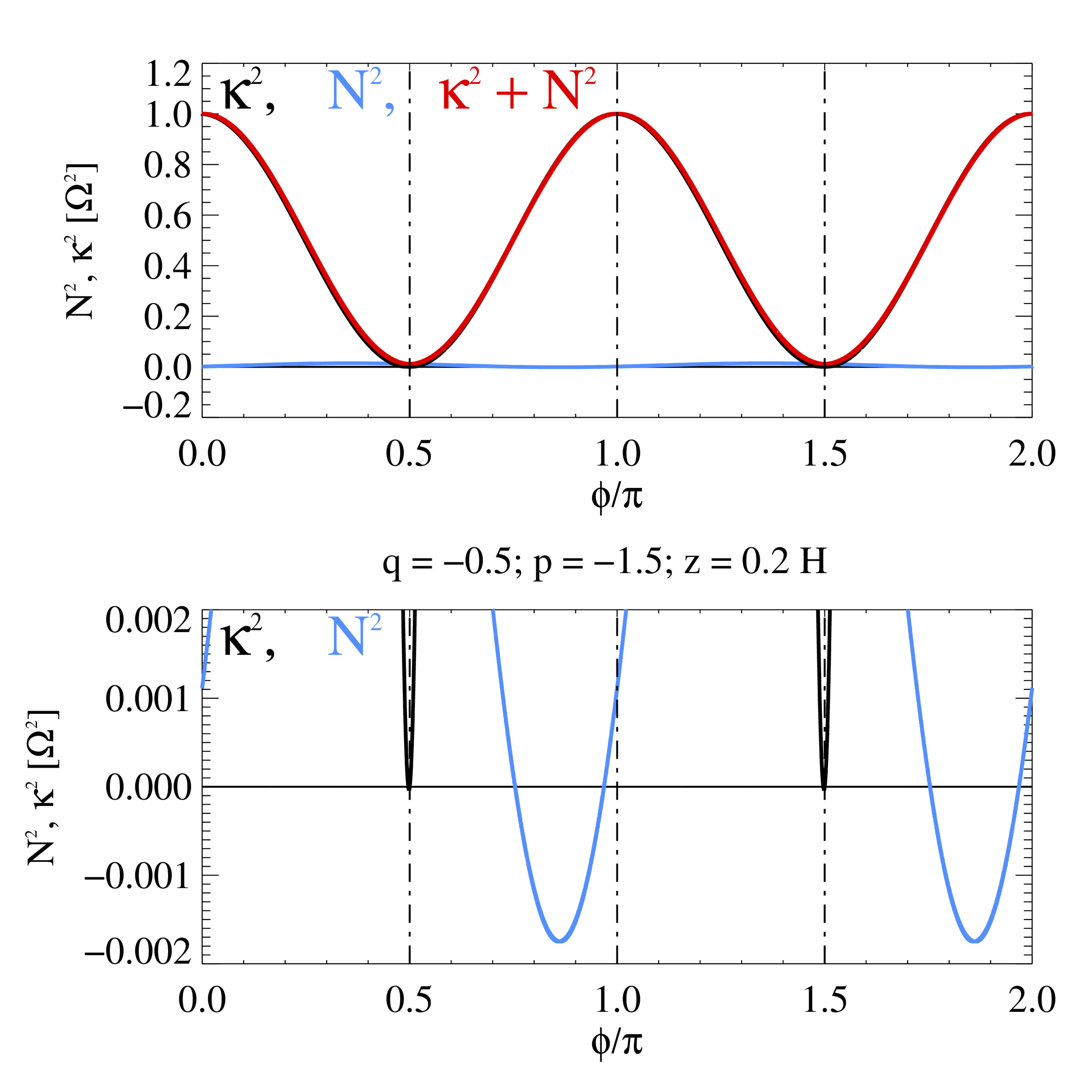}{0.5\textwidth}{(b): z = 0.2H}}
    \caption{Visualization of the Solberg-Høiland criterion for a vertically isothermal disk atmospheres with a radial temperature gradient of $q=-0.5$ and a radial midplane density gradient of $p=-1.5$. This leads to radially increasing specific entropy gradient in the midplane. We plot the directional \BVF $N^2_\mathbf{k}$ (Eq.\ \eqref{eq:BVTK}) and oscillation frequency $\kappa^2_\mathbf{k}$ (Eq.\ \eqref{eq:2dKAP}) as function of direction angle of the perturbation $\phi = \tan^{-1}(\boldsymbol{a}_R/\boldsymbol{a}_z) = -\tan^{-1}(k_R/k_z)$. On the left we show the situation of the disk at one pressure scale height $z=H$. The upper plot shows the overall sinusoidal like variation of the frequencies with direction $\phi$ with buoyancy $N^2_\mathbf{k}$ being stable (positive) for perturbations in the vertical direction (upward $\phi = \onehalf \pi$ and downward $\phi = 3\onehalf \pi$). The epicyclic frequency $\kappa^2_\mathbf{k}$ is maximal parallel to the midplane $\phi = 0$ and $\pi$. In the lower plot we emphasize the regions at which both frequencies become complex, i.e.\ the square is negative: 
    $\kappa^2_\mathbf{k}$ is negative for a narrow range of angles at $\phi < \onehalf \pi$ and $\phi < 1 \onehalf \pi$ and $N^2_\mathbf{k}$ is negative for approximately the radial direction $\phi = 0, \pi$.
    Once we plot the sum of the frequencies (red curve) we find the combined stratification to be always stable $\kappa^2_\mathbf{k}+N^2_\mathbf{k} > 0$, as discussed in the Solberg-Høiland criterion (see Section 3.1). For the thermal instabilities on the other hand it is sufficient that the individual frequencies are complex (see Section 3.2). 
    The plots on the right show the same behavior for a region closer to the mid plane $z = 0.2 H$ at which the radial \BVF $N_R^2$ is positive. But for a range of intermediate angles one finds unstable stratification. Thus $N_R^2 > 0$ and $N_z^2 > 0$ is not a sufficient criterion for stability. We added the plots in the lower row to highlight the directions in which the square of either $N^2$ or $\kappa^2$ is negative.} 
        \label{Fig:NKofZ}
\end{figure*}

\section{Conditions for Stability}
First we discuss the conditions for stability without thermal relaxation as in \cite{Ruediger2002}, then we apply instantaneous cooling as in \cite{Goldreich1967} and \cite{Fricke1968}, which is the driving mechanism for the VSI phenomenon, and finally we discuss the case for finite cooling times.
\subsection{No thermal relaxation: dynamic instabilities}
Equation \eqref{eq:n3} leads back to the Solberg-Høiland criterion for dynamical stability if one sets $\epsilon = 0$, or equivalently $\tau = \infty$, which is equivalent to the dispersion relation as in \cite{Ruediger2002}, yet without sound waves
\begin{eqnarray}
\omega ^2  = \left[N_\mathbf{k}^2 + \kappa_\mathbf{k}^2\right].
\end{eqnarray}
The $k_R/k_z$ dependence is absorbed in the expressions for $N_\mathbf{k}^2$ and $\kappa_\mathbf{k}^2$.
The stability condition now requires that there is no $\mathbf{k}$ for which the expression is negative
\begin{eqnarray}
\min\left[N_\mathbf{k}^2 + \kappa_\mathbf{k}^2\right]_\mathbf{k} \geq 0
\end{eqnarray}
We can express $N_\mathbf{k}^2 + \kappa_\mathbf{k}^2$ (see Eq.\ \eqref{eq:BVTK} and Eq.\ \eqref{eq:2dKAP}) as a quadratic function of $k_z/k_R$ \citep{Ruediger2002}. The condition that the quadratic and constant terms are larger than zero leads to the first Solberg-Høiland criterion for stability:
\begin{eqnarray}
 N_R^2 + N_z^2 + \kappa_R^2 > 0, \,\,\, \,\,\,\textbf{Solberg-Høiland Criterion I}
\end{eqnarray}
and the condition that there is no root for any $k_z/k_R$, i.e. the discriminant has to be negative, gives:
\begin{eqnarray}
a_z \left(\kappa_R^2 s_z - \kappa_z^2 s_R \right) < 0.
\,\,\, \,\,\,\textbf{Solberg-Høiland Criterion II}
\end{eqnarray}
We can now use an actual disk atmosphere to determine the local stability properties (See appendix \ref{appendix1}). As example we choose a disk which is radially convective stable in the midplane with density gradient of $p = -1.5$ and a temperature gradient of $q=-0.5$. 
In Figure \ref{Fig:NKofZ}, we plot $N_\mathbf{k}^2 + \kappa_\mathbf{k}^2$ as a function of direction $\mathbf{k}$, respectively the angle expressed by $\phi = \tan^{-1} (k_R / k_z)$ for two different heights above the midplane: $z=0.2H$ and $z=H$. We find that for the disk parameters chosen this expression is never negative, indicating dynamical stability for disks.
But note that for certain directions either $N_\mathbf{k}^2 $ or $ \kappa_\mathbf{k}^2$ can obtain negative values which will be important in the following section.

\subsection{Instantaneous cooling: GSF and VSI}
For $\tau = 0$ the dispersion relation predicts growth rates $\Gamma$ of:
\begin{eqnarray}
\Gamma ^2  = - \kappa_\mathbf{k}^2.
\end{eqnarray}
Looking for the optimum growth rate wavenumber ratio for GSF, we have to find the minimum for $\kappa_\mathbf{k}^2$
by calculating 
\begin{equation}
\frac{\mathrm{d}}{\mathrm{d} \frac{k_z}{k_R}} \kappa^2_\mathbf{k} = 0,
\end{equation}
which leads to
\begin{equation}
\left(\frac{k_z}{k_R}\right)_- = \frac{1}{2}\frac{\kappa_z^2}{\kappa_R^2}.
\end{equation}
As a result, we can define a minimum $\kappa_-^2$ for this wave number ratio
\begin{equation}
\kappa_-^2 = -\frac{1}{4}\frac{\kappa_z^4}{\kappa_R^2} \approx -\frac{1}{4}\frac{\kappa_z^4}{\Omega^2}.
\end{equation}
The resulting local growth rate for GSF under very fast cooling is
\begin{equation}
    \Gamma_\mathrm{GSF} = \frac{1}{2}\frac{|\kappa_z^2|}{\Omega} = \frac{|q|}{2} \frac{|z|}{R} \Omega,
\end{equation}
derived in a similar but not identical fashion as in \citep{Urpin1998}, who did not calculate the extremum wavenumber ratio, but estimated an optimum value for 
\begin{equation}
\frac{k_z}{k_R} = \frac{\kappa_z^2}{\kappa_R^2},
\end{equation}
thus our growth rates differ by a factor of two.

Work on the VSI as in \citet{Nelson+2016} removes the $z$ dependence of growth rate by considering certain global modes, but the VSI growth rate \citep{Lesur2023}:
\begin{equation}
    \Gamma_\mathrm{VSI} \approx |q| \frac{H}{R} \Omega,
\end{equation}
is essentially the GSF growth rate at $z = 2 H$\footnote{Note that \citet{Lin2015} compares the VSI growth rate to the approximate result in \citet{Urpin2003}}.


\subsection{Finite thermal relaxation: COS and GSF}
As soon as one includes thermal relaxation, the Solberg-Høiland criterion is no longer sufficient for stability. Our third-order dispersion relation can be written as the following polynomial \citep{Tassoul2000}:
\begin{eqnarray}
\Gamma^3  + \Gamma ^2  \frac{1}{\gamma  \tau} + \Gamma A + \frac{1}{\gamma  \tau } B = 0.
\end{eqnarray}
The term $\frac{1}{\gamma  \tau}$ is always positive, thus
the terms $A$ and $B$ define the possible solutions. The dispersion relation has three roots, of which two are oscillatory $\Gamma_{1,2} = a_\mathrm{T} \pm i \omega$ and one is non-oscillatory $\Gamma_3 = b_\mathrm{T}$.
According to the Routh-Hurwitz criterion \citep{Tassoul2000,Balbus2001}, the terms for the growth rates $a_\mathrm{T}$ and $b_\mathrm{T}$ are negative if and only if
\begin{eqnarray}
B > 0 \,\, \mathrm{and} \,\, A - B > 0.
\end{eqnarray}
The two conditions in our notation are
\begin{eqnarray}
\kappa_\mathbf{k}^2 > 0 \,\, \mathrm{and} \,\, N_\mathbf{k}^2 > 0.
\end{eqnarray}
The violation of the first condition leads to GSF (and VSI) as in the instantaneous cooling case (see previous subsection). This is a thermal instability in the terms of \citet{Tassoul2000}. 
To be stable to thermal instability one needs to fulfill:
\begin{eqnarray}
  \kappa_z^2 = 0      \textbf{\,\,\, \,\,\,GSF Criterion.}
\end{eqnarray}
Thermal overstability occurs for $N_\mathbf{k}^2 < 0$.
In the same fashion as for the Solberg-Høiland criterion for stability we have to make sure that the quadratic expression for $N_\mathbf{k}^2$ is strictly positive for all $\mathbf{k}$.

\begin{figure*}
    \centering
    \gridline{\fig{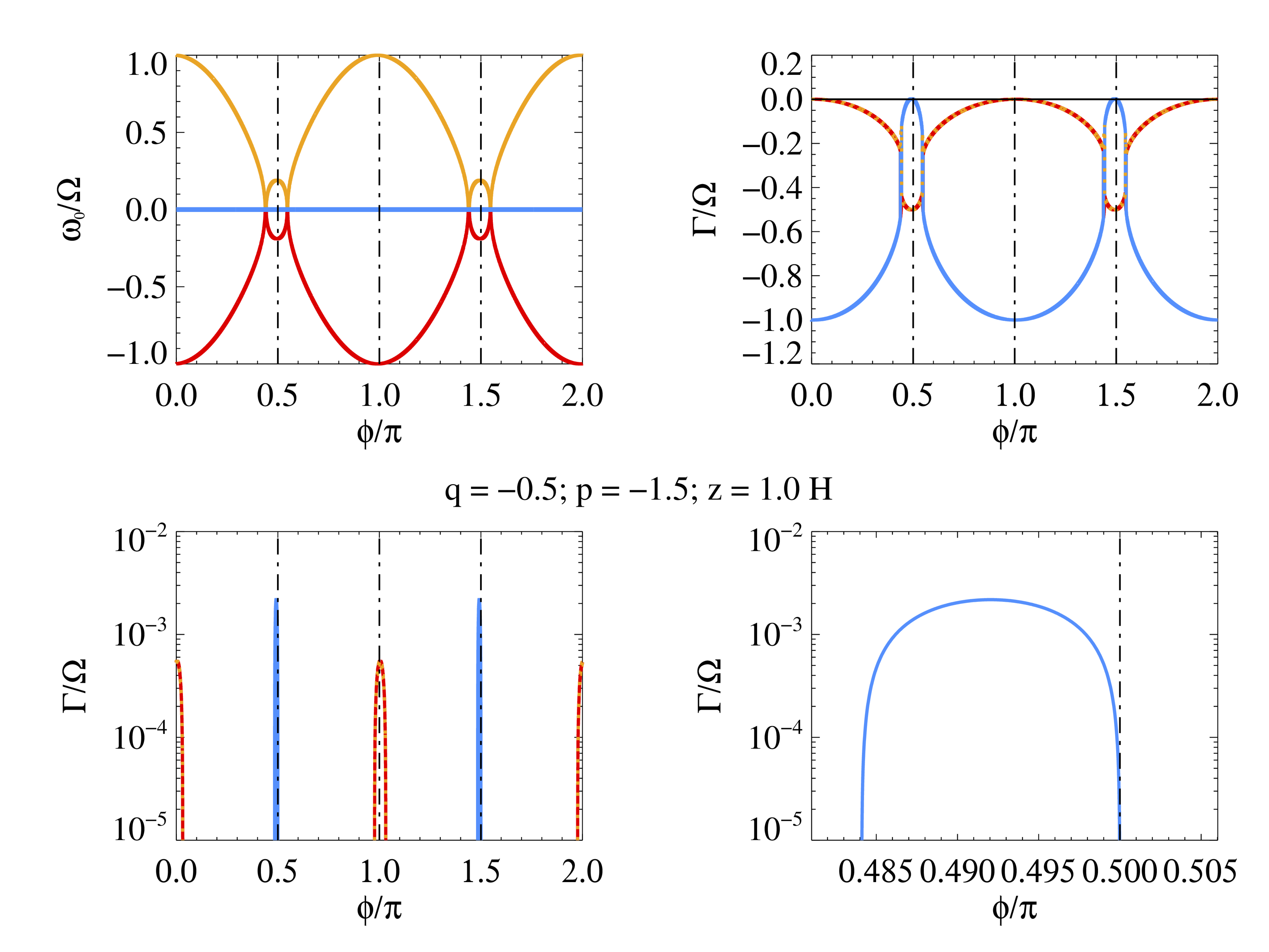}{0.5\textwidth}{(a): z = H}\fig{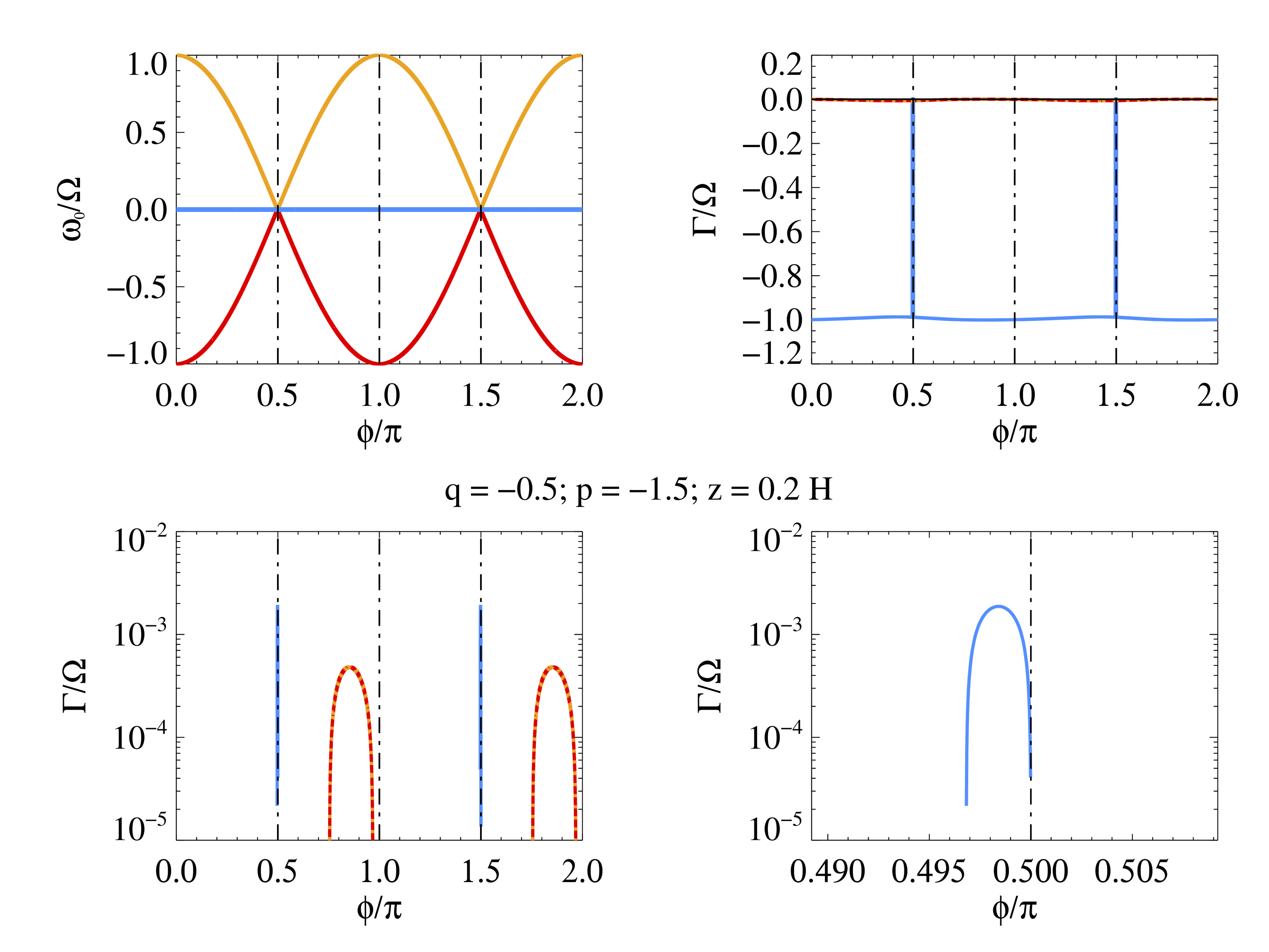}{0.5\textwidth}{(b): z = 0.2H}}
    \caption{Numerical solution for the dispersion relation (Eq.\ \eqref{Eq:FullDispOmega}) for a disk with a radial temperature gradient of $q = -1$, $p=-1.5$ and a cooling time of $\tau = 1 / \Omega \gamma$ for two different locations of the disk (a) $z = H$ and (b) $z = 0.2 H$. As expected, we find two complex roots and one imaginary one. 
    The imaginary part gives the growth rates, if larger than zero.
    Depending on the direction of the perturbation $\phi$ we find stable and unstable solutions. The non-oscillatory solutions are the strongest for the vertical direction and we identify them as the GSF modes. The modes oscillating with the Keplerian frequency (epicyclic frequency) are the COS modes. They are the fastest-growing modes for the radial direction at $z = H$. For regions closer to the midplane, the oscillations' radial direction is actually damped, yet in an inclined direction there is still growth for the overstable modes. Note that the growth rates for both heights are very similar.\label{Fig:NgrateQ1T1}} 
\end{figure*}


To check if $N_\mathbf{k}^2$ is negative for any $k$ we write down the full expression as a quadratic function of $k_R/k_z$
\begin{eqnarray}
\frac{k^2}{k_z^2} N^2_\mathbf{k}  &=&  \frac{k_R^2}{k_z^2} N^2_z + N^2_R + \frac{c^2}{\gamma}\frac{k_R}{k_z} (b_R s_z + b_z s_R)
  \label{eq:2dBV0}
\end{eqnarray}
Pure vertical modes $k_z = 0$ would only be unstable if the disk was vertically convective unstable, which is a case that we not consider. This case would already lead to a violation of the Solberg-Høiland criterion. So we can exclude this case and divide by $k_z^2$, which is also strictly positive.
This leads to a quadratic equation of the type:
\begin{align}
\frac{k_R^2}{k_z^2} A  + \frac{k_R}{k_z} B + C.
  \label{eq:2dBVa}
\end{align}
To be always positive, the sum of the constant $C$ and quadratic $A$ term have to be positive in this quadratic equation, thus we get the first stability criterion:
\begin{eqnarray}
   N^2_z + N^2_R > 0. \textbf{\,\,\,\,\,\, COS Criterion I}
  \label{eq:2dBVX}
\end{eqnarray}
At the midplane and in the case of no or adiabatic vertical stratification, the radial \BVF alone determines stability, as in \citet{Klahr2014}.
But in the general case of vertical stratification, Equation (\eqref{eq:2dBVX}) alone cannot ensure stability. The discriminant $B^2 - 4 A C < 0$ also has to be negative:
\begin{eqnarray}
b_R^2 s_z^2 + 2 b_R s_z b_z s_R + b_z^2 s_R^2 - 4 b_z s_z b_R s_R < 0.
  \label{eq:2dBVY3}
\end{eqnarray}
With a little algebra using Eq.\ (\eqref{Eq.kappa_z}) we find:
\begin{eqnarray}
b_R^2 s_z^2 - 2 b_R s_z b_z s_R + b_z^2 s_R^2 = (b_R s_z - b_z s_R)^2 = \frac{\gamma^2}{c^4} (\kappa_z^2)^2 < 0.
  \label{eq:2dBVY4}
\end{eqnarray}
which as a second condition for stability can be expressed as
\begin{eqnarray}
 \kappa_z^4 = 0, \textbf{\,\,\,\,\,\, equiv.  \,\,\,\,\,\,} \mathbf{\nabla}P \times \mathbf{\nabla}\rho = 0 \textbf{\,\,\,\,\,\, COS Criterion II}
  \label{eq:2dBVY4B}
\end{eqnarray}
because $\kappa_z^4$ cannot be negative. Therefore, only a disk without vertical shear does not show thermal overstabilty.
This second condition is the same condition as for GSF (and VSI), i.e.\ baroclinic disks with vertical shear are never stable for thermal overstability. Thus all atmospheres unstable to GSF (and VSI) are also unstable to COS. \response{These stability criteria and their derivation are identical to Eq.\ (62), Eq.\ (63) and Eq.\ (64) in \citet{Menou2004}.}

The case for a radial temperature distribution in a disk of $q = -0.5$ as shown in Fig.\ \ref{Fig:NKofZ} is thus a typical situation in a disk. 
If $\kappa^2_k$ is negative in some direction, then there must be also a $N_-^2 < 0$ in some other direction. How exactly these quantities are related, will be discussed in the following section.

As both conditions for COS have to be fulfilled for stability, locations in a disk which are not unstable to VSI, as for instance the midplane with sufficient entropy gradients, can be unstable to COS because of a sufficient radial entropy gradient.
 This condition reduces the first COS criterion for stability to
 \begin{eqnarray}
   N^2_R > 0. \textbf{\,\,\,\,\,\, COS Criterion I}.
  \label{eq:2dBVXM}
\end{eqnarray}
For a barotropic atmosphere with vertical stratification, this criterion would also be sufficient, for in that case $N_R^2$ and $N_z^2$ must have the same sign. Yet, for the unstable case in a bartropic disk, $N_R^2 < 0$ would imply $N_z^2<0$ and one is not only potentially COS unstable, but can already violate the second \SH criteria. One might then have vertical thermal convection for instance \citep{Tassoul2000} in the absence of cooling.
With sufficient cooling on the other hand convection would be damped. 
\begin{figure*}
    \centering
       \gridline{\fig{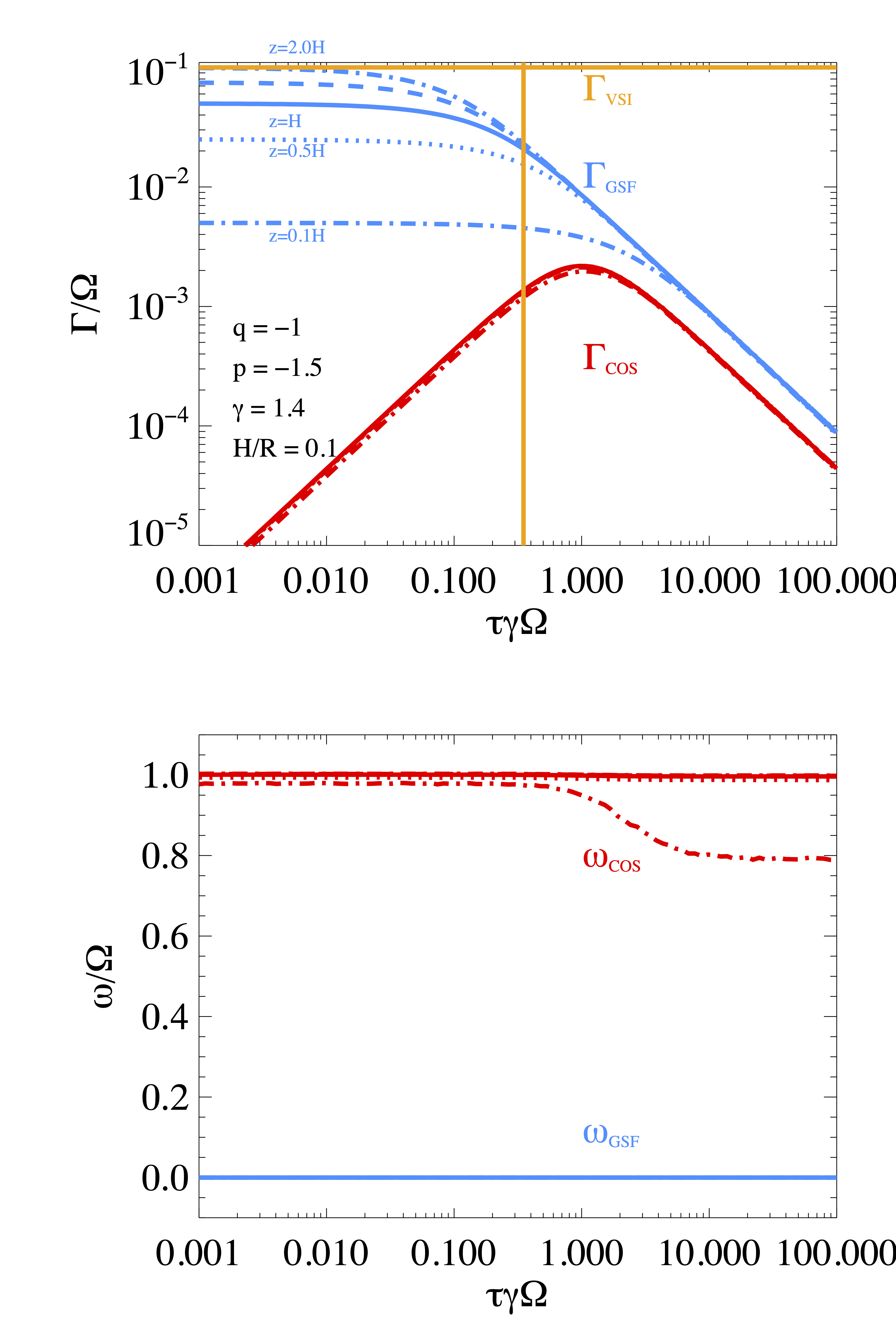}{0.43\textwidth}{(a): q = -1}\fig{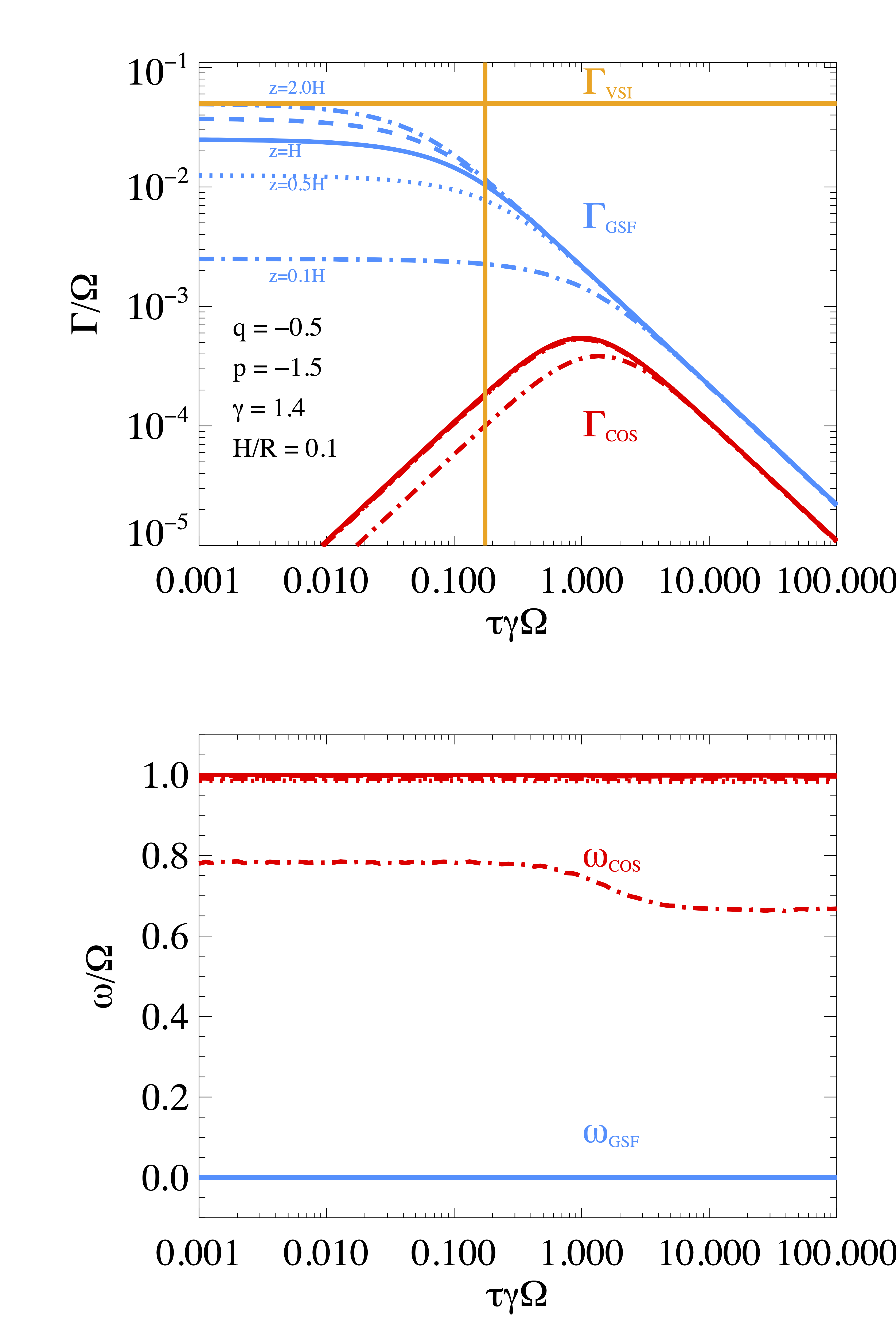}{0.43\textwidth}{(b): q = -0.5}}
    \caption{Growth rates $\Gamma$ and frequency $\omega$ for COS and GSF for $p = -1.5$ as a function of cooling time $\tau$ for various heights above the midplane. We plot the numerical solution of the dispersion relation (Eq.\eqref{Eq:FullDispOmega}). (a): $q = -1$ and (b): $q = -0.5$. In ascending order $z =  0.1H, 0.5H, H, 1.5 H, 2 H$ for a disk with $H/R = 0.1$ and the adiabatic index of $\gamma = 1.4$. The thickest blue line corresponds to $z=H$. The strength of COS is independent of height, which makes the red line almost indistinguishable. The oscillation frequency for GSF is always zero. For COS it is mostly $\omega = \Omega$, and only when close to the midplane does buoyancy, and the fact that the most unstable modes are not radial, modify the frequency slightly. The yellow vertical line indicates the critical cooling time for standard VSI $\tau_\mathrm{c}$ and the yellow horizontal line is the associated zero-cooling-time growth rate, as found in the literature \citep{Nelson2013}.
    } 
        \label{Fig:TAUGRATE4Q1}
        
\end{figure*}
\section{Growth rates as function of cooling time}
We can now solve the dispersion relation numerically as a function of the wavenumber pair $k_x, k_y$ for various locations in the disk atmosphere $z,R$, for the local gradients in pressure and entropy respectively. If we assume a vertically isothermal atmosphere, then the disk aspect ratio $h$ at radius $R$ together with the midplane radial density gradient $p$ and temperature gradient $q$ (see definitions in Table 1) describe an approximate Gaussian density distribution (see for instance \citet{Nelson2013})
\begin{equation}
    \rho = \rho_0 \left(\frac{R}{R_0}\right)^p  \exp\left(-\frac{z^2}{2 H^2}\right),
\end{equation}
from which you can calculate all the necessary entropy and pressure gradients $s_R, s_z, b_R$ and $b_z$ that we describe in Appendix A.

For each parameter set, the dispersion relation has three unique solutions for the complex $\omega = \omega_0 + i \Gamma$. Here, $\omega_0$ is the real part of the root and $\Gamma$ is the complex part. When $\Gamma$ is positive, we find a growing mode, when it is negative the mode decays.
In Figure \ref{Fig:NgrateQ1T1}, we show the growth rates $\Gamma > 0$ for instabilities in a disk atmosphere with a radial temperature gradient of $q=-1$ and a cooling time of $\tau^* = 1$ for a range of wavenumber pairs, or, more precisely, the direction of the considered perturbation. 
In the case of a convective stable midplane with $q=-0.5$ for $p=-1.5$, both vertical and radial \BVFs are positive, but there are nevertheless inclined convective unstable modes ($k_R / k_z \approx 0.2$) at $z = 0.2H$ as can be seen in Figure \ref{Fig:NgrateQ1T1}. At a scale height of $z=H$ there are radial convective modes with $k_R=0$. For all heights we also find GSF modes with $k_R < 0$, i.e.\ not in the vertical direction, but half-way to the contour of constant specific angular momentum. 

We can classify the numerical solutions of the dispersion relation: if there is no real oscillatory part, then it is a GSF mode, and if it oscillates we classify it as COS.

In Figure \ref{Fig:TAUGRATE4Q1}, we plot the identified growth rates for GSF: $\Gamma_\mathrm{GSF}$ and COS: $\Gamma_\mathrm{COS}$ as function of cooling time for various heights in the disk with either radially stable midplane stratification ($h=0.1$, $p=-1.5$, $q=-0.5$) or an unstable midplane ($h=0.1$, $p=-1.5$, $q=-1.0$). The upper plot gives the growth rates for COS (red), for GSF (blue) as well as VSI (yellow horizontal line). We additionally plot the critical cooling time for VSI \citep{Lin2015} as vertical yellow line. The lower plots give the associated oscillation frequencies for the same modes, which we used for the GSF versus COS classification. $\omega_\mathrm{GSF}$ is always zero as expected \response{from the local analysis, which distinguishes these modes from vertically global VSI modes with a finite oscillation frequency as discussed in \citet{Lin2015}.} The frequency for COS $\omega_\mathrm{COS}$ typically is close to the epicyclic frequency. Only within a narrow range near the midplane, where the COS unstable modes deviate from being purely radial, does the oscillation frequency decrease, as anticipated from the projection of the epicyclic frequency and the radial stable stratification for $q=-0.5$, due to: $\omega_0^2 \approx \kappa_R^2 + N_R^2$.

\subsection{Analytic expressions for growth rates}\label{SS:AnalyticExpressions}
\citet{Urpin2003} derived growth rates for both thermal instabilities as a function of cooling rate and $N^2_- = \min(\omega_g)$ and $\kappa^2_- = \min(Q^2)$, but did not explicitly determine the corresponding values for $N^2_-$, assuming that it is always positive. Thus we first determine the actual values of $N^2_-$ for an accretion disk.
\begin{figure*}
    \centering
    \gridline{\fig{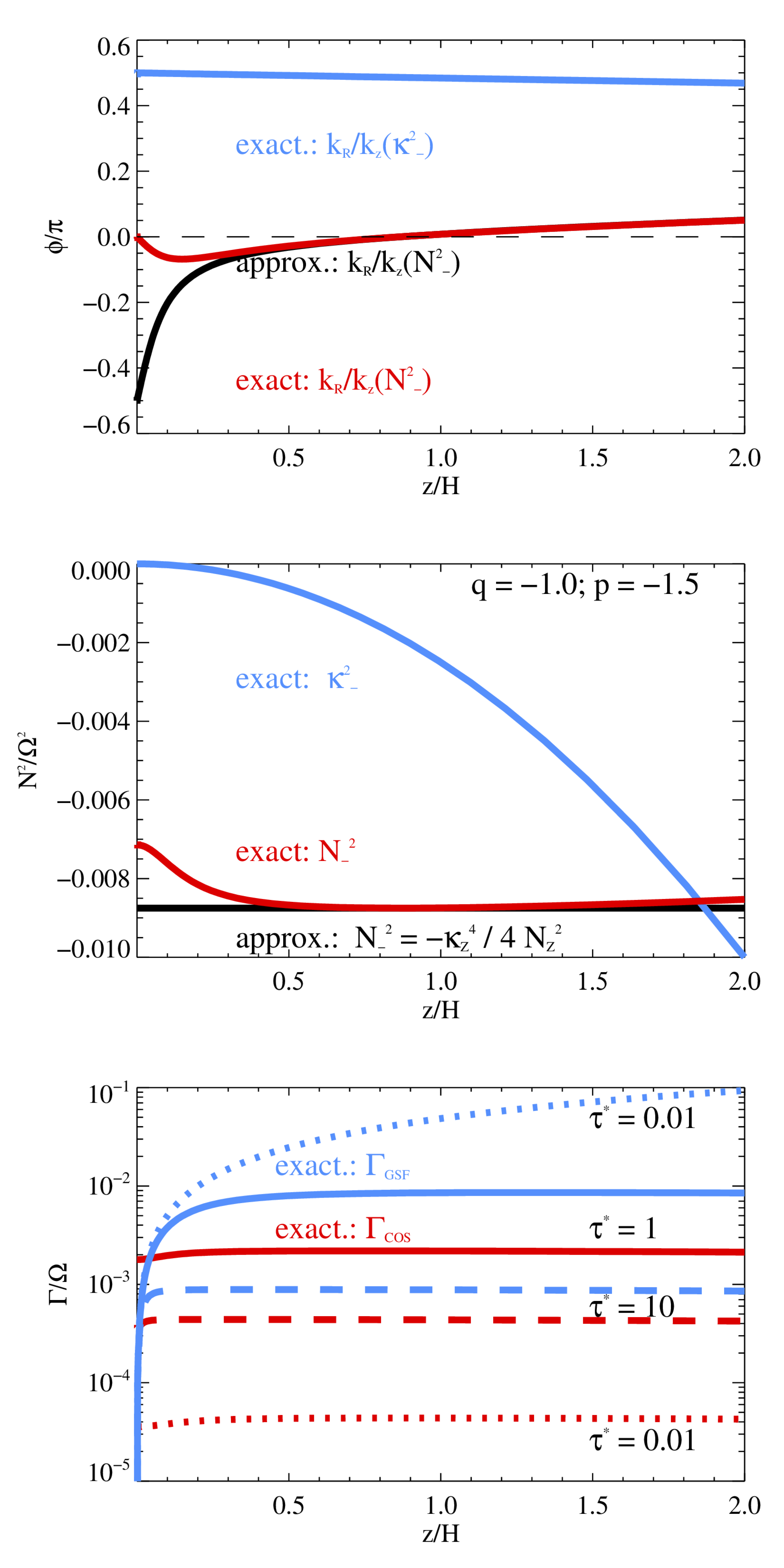}{0.5\textwidth}{(a): q = -1: unstable midplane $N_R^2 < 0$}\fig{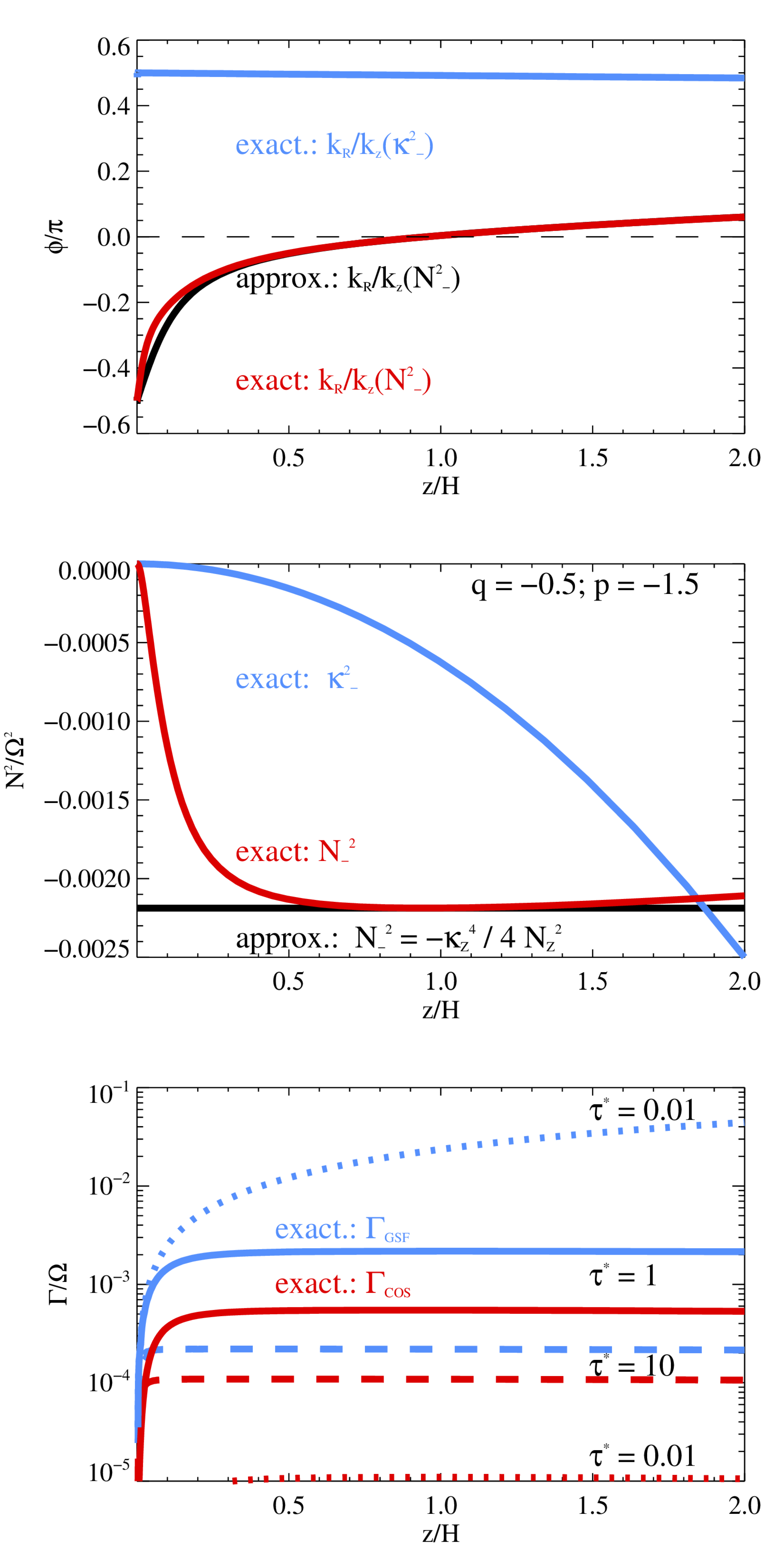}{0.5\textwidth}{(b): q = -0.5 stable midplane $N_R^2 > 0$}}
    \caption{Upper row: angle $\phi = -\tan^{-1}{k_R / k_z}$ for maximal buoyant instability (red line), i.e.\ the minimum of $N^2_-$ modes for wave-number $\mathbf{k_-}$ (Eq.\ \eqref{Eq:k_Rdk_z}) as a function of height above midplane for (a) a radial unstable midplane with $q=-1$ and (b) a stable midplane $q=-0.5$. $\phi = 0 $ indicates radial modes. The black line is the approximated value obtained from Eq.\ (\eqref{Eq:k_Rdk_za}). We also plot the angle for the strongest GSF mode $\kappa^2_-$ for as a blue line ($\phi = 0.5 \pi$ indicates the vertical direction). Middle row: We also plot exact (Red) and approximate values (Black) for $N^2_-$ based on $\mathbf{k_-}$. For $q=-1$ the approximation $N^2_- N^2_z \approx - \kappa_-^2 \Omega^2 $ based on $\kappa_-^2$ (blue line) gives a good value for all considered heights. For a stable midplane $q=-0.5$ with $N_R^2 >0 $ we still find our approximation to fit well above about $z \approx 0.3 H$. Lower row: We also plot resulting growth-rates for GSF (blue) and COS (red) at three different thermal relaxation times: $\tau^* = 10$ (dashed lines), $1$ (solid lines) and $0.01$ (dotted lines).  \label{Fig:N2MKQ1Q05}} 
\end{figure*}

The largest growth rates for COS will occur for perturbations in the direction where $N_\mathbf{k}^2$ is the lowest (and negative). 
So we define $N^2_+$ as the largest possible buoyancy frequency and $N^2_-$ as the smallest for a certain angle of the wave vector $\phi = -\tan^{-1}(k_R/k_z)$.
For a given $k^2 = 1$ and $k_R = \sin{(\phi)}$ and $k_z = -\cos{(\phi)}$ we find:
\begin{eqnarray}
\frac{d N^2_\mathbf{k}}{c^2 d \phi}  = \partial_\phi \left(\left(-\cos{(\phi)} b_R  -\sin{(\phi)} b_z \right) \left(\sin{(\phi)} s_z  + \cos{(\phi)} s_R   \right)\right) = 0,
\end{eqnarray}
which is
\begin{eqnarray}
 (\sin(\phi) b_R  - \cos(\phi) b_z ) \left(\sin(\phi) s_z  + \cos(\phi) s_R \right) + (-\cos(\phi) b_R -\sin(\phi) b_z ) \left(\cos(\phi) s_z  - \sin(\phi) s_R   \right) = 0.
\end{eqnarray}
Rearranging the terms leads to:
\begin{eqnarray}
 (\sin^2(\phi) - \cos^2(\phi)) (b_R s_z + b_z s_R) + 2 \sin(\phi) \cos(\phi) \left(b_R s_R - s_z b_z \right) = 0.
\end{eqnarray}
Vertical modes with $\cos(\phi) = 0$ can be ignored and we also define $b_R s_z + b_z s_R\ne 0$ so we can divide by these terms,
which leads to a quadratic equation for the $\tan$:
\begin{eqnarray}
 \tan(\phi)^2 + 2 \tan(\phi) \frac{b_R s_R - s_z b_z}{b_R s_z + b_z s_R} - 1 = 0,
\end{eqnarray}
as a way to find the extrema of $N^2$ as function of $k_R/k_z$. Thus, the extremal values of $k_R/k_z$ are
\begin{eqnarray}
\left(\frac{k_R}{k_z}\right)_\pm = - \tan(\phi)_\pm = \frac{b_R s_R - b_z s_z}{b_R s_z + s_R b_z} \pm \sqrt{\left(\frac{b_R s_R - b_z s_z}{b_R s_z + s_R b_z}\right)^2 + 1}.
\label{Eq:k_Rdk_z}
\end{eqnarray}
It is immediately clear that both solutions are orthogonal vectors:
\begin{eqnarray}
\left(\frac{k_R}{k_z}\right)_+  \left(\frac{k_R}{k_z}\right)_-= - \left(\frac{k_R}{k_z}\right)_+  \left(\frac{k_z}{k_R}\right)_+ = -1.
\label{Eq:k_Rdk_z_ident}
\end{eqnarray}

We plot the values for $\phi = - \tan^{-1}(k_R/k_z)$ in Fig.\ (\ref{Fig:N2MKQ1Q05}) as a function of height.
We also add the values for $N^2_-$ calculated for this 
angle. Surprisingly, the absolute value of $N_-^2$ does not strongly depend on height once outside the midplane, especially considering that $N^2_z$ is a quadratic function with height. 
\begin{equation}
    N^2_z = \frac{\gamma-1}{\gamma} \frac{z^2}{H^2} \Omega^2.
\end{equation}
Plugging the full expression for $\frac{k_R}{k_z}$ into the expression of $N^2$ does not lead to much insight, so we approximate $\frac{k_R}{k_z}$ by defining it to be very small, implying that $k_z \ne 0$ and divide the expression for the directional \BVF by $k_z^2$.
\begin{eqnarray}
N^2_\mathbf{k}\frac{k_R^2 + k_z^2}{k_z^2} \approx N^2_\mathbf{k}.
\end{eqnarray}
Then we can take the derivative of $N^2_\mathbf{k}$ with respect to $k_R/k_z$.
\begin{eqnarray}
 \frac{\partial (\frac{k_R^2}{k_z^2} N^2_z + N^2_R + \frac{k_R}{k_z} \frac{c^2}{\gamma}(b_R s_z + b_z s_R))}{\partial (k_R/k_z)} = 2 \frac{k_R}{k_z} N^2_z+ (b_R s_z + b_z s_R) = 0,
  \label{eq:2dBV718}
\end{eqnarray}
Thus the extremum in $N^2_\mathbf{k}$ is approximately at:
\begin{eqnarray}
 \frac{k_R}{k_z} = - \frac{b_R s_z + b_z s_R}{2 b_z s_z},
  \label{Eq:k_Rdk_za}
\end{eqnarray}
which is a first order Taylor expansion of Eq.\ \eqref{Eq:k_Rdk_z} for $b_z s_z \gg (b_R s_z + b_z s_R)$ and assuming $b_z s_z \gg b_R s_R$.
We can reintroduce this expression into Eq.\ (\eqref{eq:BVTK}):
\begin{eqnarray}
 N^2_- \approx - \frac{c^4}{\gamma^2}\frac{(b_R s_z + b_z s_R)^2}{4 N^2_z} + N^2_R = - \frac{\frac{c^4}{\gamma^2}(b_R s_z - b_z s_R)^2 + 4 N_R^2 N_z^2}{4 N^2_z} + \frac{4 N_R^2 N_z^2}{4 N_z^2} = - \frac{\kappa_z^4}{4 N^2_z}.
  \label{eq:2dBV2}
\end{eqnarray}
This results in an approximation for
the most negative value for the true height dependent $N^2_-$ as can be seen in Fig.\ \ref{Fig:N2MKQ1Q05}. In general it is a good estimate for the level of instability at least outside the midplane and much easier to be calculated.
In the midplane, where we should expect $N^2_-(z=0) = N_R^2$ as $N_z^2 = 0$, our estimate fails, \response{because here the assumption leading to Eq.\ \eqref{Eq:k_Rdk_za} is no longer fulfilled.}

The obvious dependence between vertical shear and the strength of unstable buoyancy stratification led to the assumption of a more general relation, in which the vertical \BVF $N^2_z$ is replaced by the largest stable wavelength
\begin{eqnarray}
 N^2_- = - \frac{1}{4} \frac{\kappa_z^4}{N^2_+},
 \label{eq:790}
\end{eqnarray}
for which we give a proof in the appendix. If we also use the definition of the largest and smallest projected oscillatory frequency $\kappa^2$ we find
\begin{eqnarray}
 N^2_- N^2_+ = \kappa_-^2 \kappa_+^2,
 \label{eq:790b}
\end{eqnarray}
as a general property of baroclinic atmospheres.
This elucidates our criterion for thermal overstability COS once more: To ensure that $N^2_- \geq 0$ one needs $\kappa_z^4 = 0$.



\begin{figure*}
    \centering
       \gridline{\fig{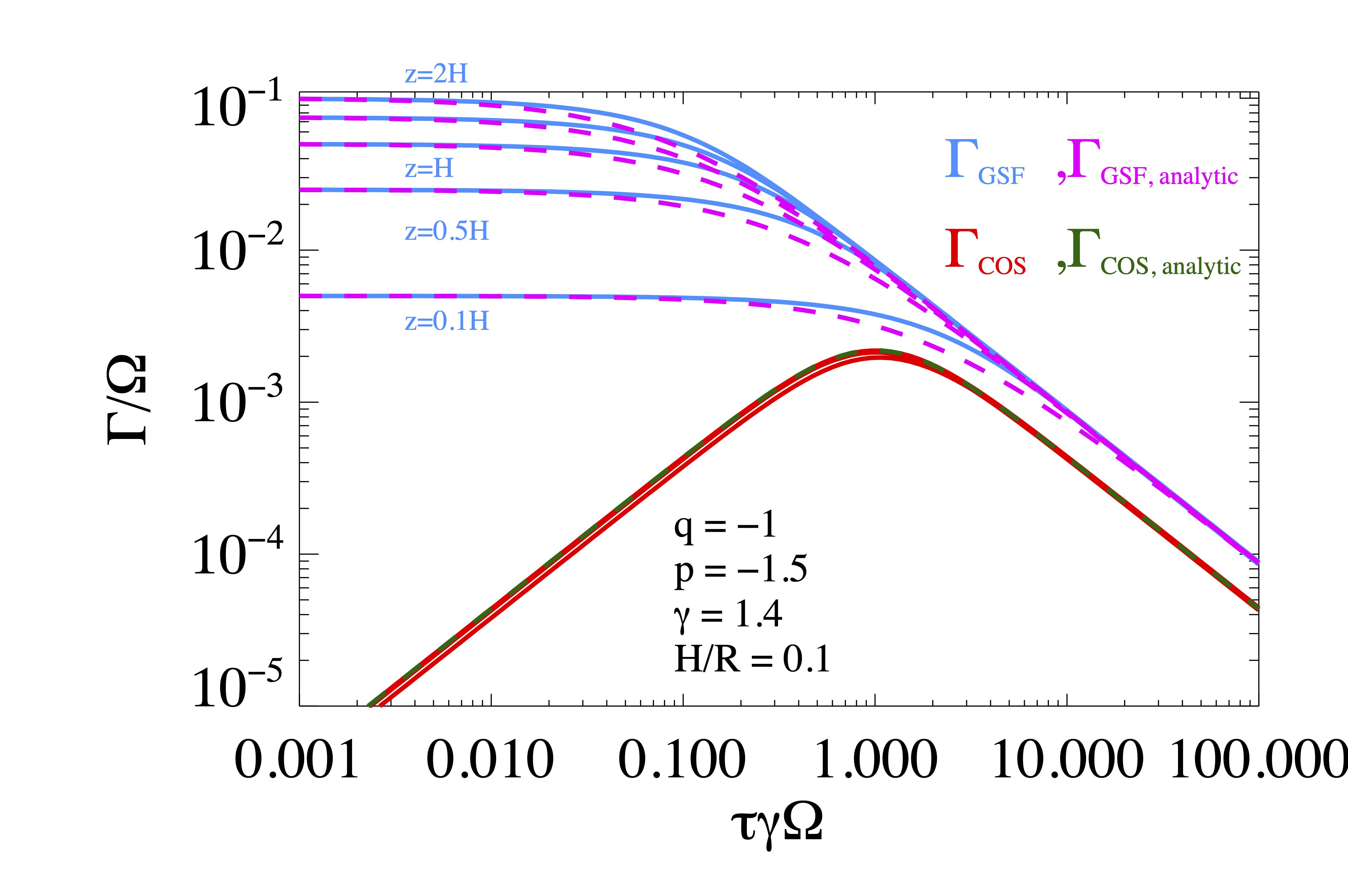}{0.5\textwidth}{(a): q = -1}\fig{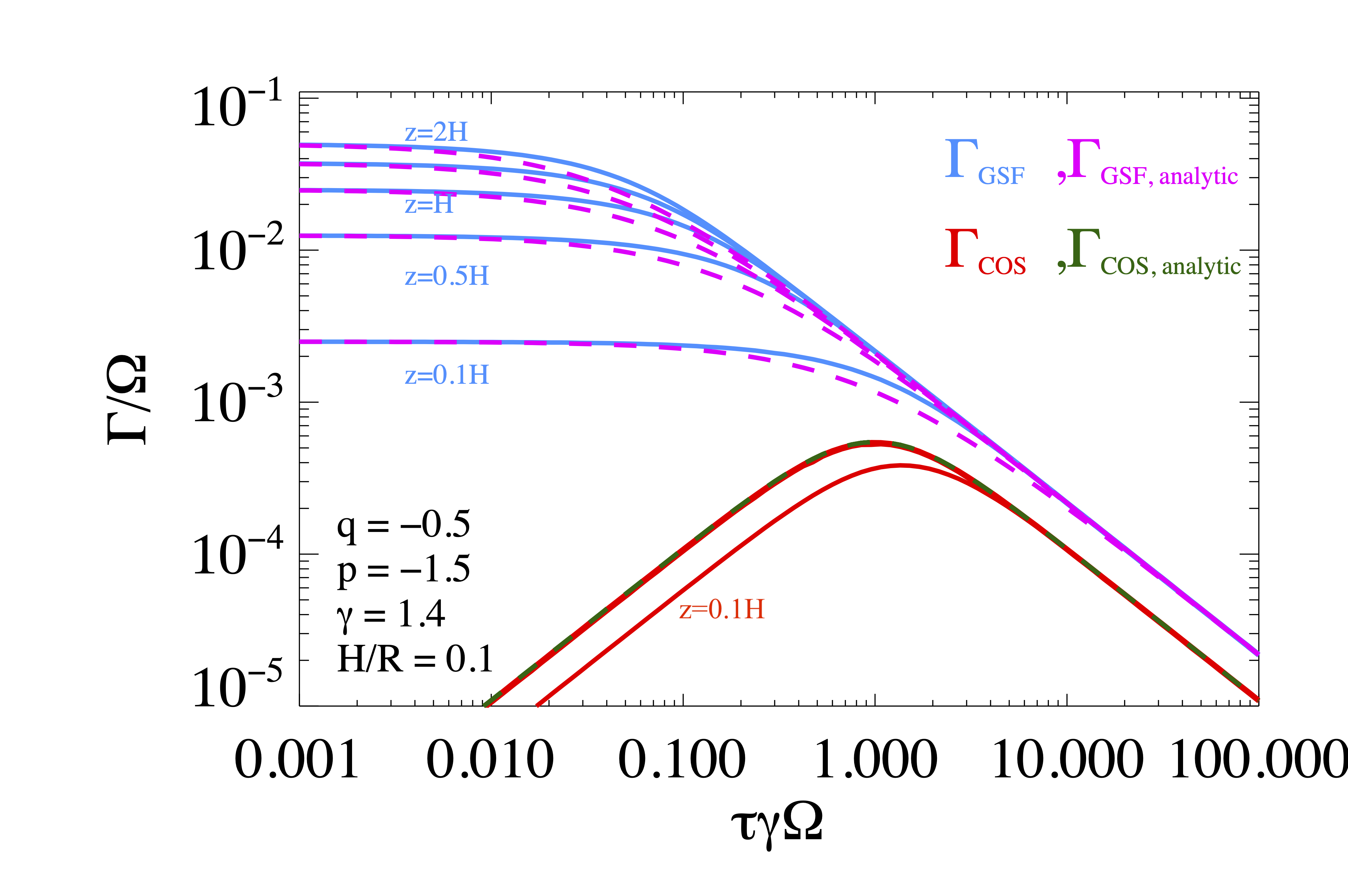}{0.5\textwidth}{(b): q = -0.5}}
    \caption{Comparison of growth rates $\Gamma$ between the numerical solution of the dispersion relation (Eq.\ \eqref{Eq:FullDispOmega}) and the approximate expressions for GSF (Eq.\ \eqref{Eq:GAMMA_GSF}) and COS (Eq.\ \eqref{Eq:GAMMA_COS}). The overall fit is very good. Deviations in the GSF rates occur for cooling times around the critical cooling for the transition from quasi instantaneous cooling to finite cooling times. And for COS we overestimate the growth rates in layers close to the midplane for $q=-0.5$ (true value for $z=0.1 H$ is the only curve not overlapping with estimated and true values for other heights.).
    } 
        \label{Fig:TAUGRATE4AQ1}
\end{figure*}

\subsubsection{COS}
Our general dispersion relation is very similar to the one for the unstratified case \citet{Klahr2014,Lyra2014}, so we can use the same strategy to derive the growth rates $\Gamma$ for convective modes. Here we uses the assumption that $\Gamma \ll \omega_0$, which we have already shown numerically in the previous section and develop the solutions around the real oscillatory frequency $\omega_0^2 = \kappa_+^2 + N_-^2 \approx \Omega^2 + N_-^2$, i.e.\ the epicyclic frequency modified by radial buoyancy. 
$\omega_0$ is the exact solution for $\tau \rightarrow \infty$ and still a good solution for finite $\tau$.
The full solution is cumbersome as noted by \citet{Urpin2003}, yet as shown in Appendix \ref{appendixCOS}, we can obtain a solution, if we drop the quadratic and cubic terms in $\Gamma$, yet keep them in $\omega_0$ and find:
\begin{equation}
    \Gamma = -\frac{1}{2}\frac{\gamma \tau N_{-}^2}{1 + \gamma^2  \tau^2  \left(N_{-}^2 +\Omega^2\right)},
\end{equation}
\response{which for the midplane $N_-^2 = N_R^2$ is identical to the unstratified case as found in \citet{Klahr2014} as Eq.\ (27) and in \citet{Lyra2014} as Eq.\ (25).}
If we use the 
approximation $|N_{-}^2| \ll \Omega^2$ and the generalized cooling time $\tau^* = \tau \gamma \Omega$ then we find the general growth rate for COS modes at a local height $z$ as
\begin{equation}
    \Gamma = -\frac{1}{2}\frac{\tau^* N_-^2}{1 + {\tau^*}^2} \Omega^{-1}. 
    \label{Eq:GAMMACOSZ}
\end{equation}
\response{\citet{Urpin2003} finds an almost identical expression (his Eq.\ (30)) for the growing oscillatory mode with complex growth rates $\gamma_{2,3}$, yet considers only the case of ${\tau^*}^2 \gg 1$,}
\begin{equation}
    \gamma_{2,3} \approx \pm i
\sqrt{Q^2 + \omega_g^2} - \frac{\omega_\chi \omega_g^2}{2 (Q^2 + \omega_g^2)},
\end{equation}
\response{where $Q^2 = \kappa_k^2$, $\omega_g^2 = N_-^2$ and $\omega_\chi = 1 / {\tau^*}$. He subsequently argues that growth is impossible, assuming $\omega_g^2$ to always be positive in "convectively stable Keplerian discs," a notion we challenge in this paper by demonstrating the contrary.}

Utilizing the characteristic value $N^2_- \approx - \kappa_z^4 / 4 N_z^2$ derived from Eqs. \eqref{eq:2dBV2} and \eqref{Eq:kappaz2} with respect to $q$ and $h$, we observe that the characteristic COS growth rate is directly given by the global temperature gradient $q$ and the local pressure scale height $h$, expressed as:
\begin{equation}
    \Gamma_\mathrm{COS} = \frac{h^2 q^2}{8} \frac{\gamma}{\gamma -1 }\frac{\tau^*}{1 + {\tau^*}^2} \Omega.
    \label{Eq:GAMMA_COS}
\end{equation}
We found that this characteristic formula aligns closely with the numerical solutions for the fastest COS growth rates obtained from the full dispersion relation (refer to Figure \ref{Fig:TAUGRATE4AQ1}). In other words, the characteristic values neatly coincide with the numerical results. 

\subsubsection{GSF}
One can also find a general solution for GSF modes with finite cooling times, if one assumes that vertical instabilities have no oscillatory solutions: $\omega = i \Gamma$, which we also demonstrated numerically. The "local" growth rates for GSF are thus recovered for $\tau \rightarrow 0$ in Equation \eqref{Eq:FullDispOmega}
and the resulting local growth rate for GSF under very fast cooling is
\begin{equation}
    \Gamma_\mathrm{GSF}(\tau \ll \tau_c) = \frac{1}{2}\frac{|\kappa_z^2|}{\Omega^2} \Omega = \frac{|q|}{2} \frac{|z|}{R} \Omega
\end{equation}
with $\tau_c = h \frac{|q|}{\gamma -1} \Omega^{-1}$ the critical cooling time for VSI \citep{Lin2015}.
If we consider finite cooling times $\gamma \tau > \frac{|\kappa_z|}{N_z^2}$ and use our knowledge that $\Gamma^2 < \kappa_-^2$ (see Fig.\ \ref{Fig:NKofZ}) then we can drop the cubed and quadratic term from the dispersion relation (for details see Appendix \ref{AppendixGSF}) and find
\begin{equation}
    \Gamma_\mathrm{GSF}(\tau \gg \tau_c) = - \frac{\kappa_-^2}{\gamma \tau \left( \kappa_-^2 + N_z^2\right)} \approx \frac{\kappa_z^4}{4 \tau \gamma \Omega^2 N_z^2} = \frac{h^2 q^2}{4} \frac{\gamma}{\gamma -1 }\frac{1}{\tau^*} \Omega,
    \label{Eq:70}
\end{equation}
as long as $N_z^2 \gg |\kappa_-^2|$. 
\response{Also this result can be found in \citet{Urpin2003} (his Eq.\ (30)) for the linear growing mode with the growth rates $\gamma_{1}$ in the regime of $\omega_\chi^{-1} = {\tau^*} \gg 1$,}
\begin{equation}
    \gamma_{1} \approx \ - \frac{\omega_\chi Q^2}{Q^2 + \omega_g^2}.
\end{equation}
\response{These growth-rates have no wavelength dependence, but as shown by \citet{Lin2015} the vertically global VSI modes can only grow in this regime for large wave-numbers $k H \ge 30$.}

We can also define a critical cooling time for GSF and find by setting $\Gamma_\mathrm{GSF}(\tau \gg \tau_c) = \Gamma_\mathrm{GSF}(\tau \ll \tau_c)$ with
\begin{equation}
    \tau_\mathrm{c,GSF}^* = \gamma \Omega \tau_\mathrm{c,GSF} = |q| \frac{h}{2} \frac{H}{|z|} \frac{\gamma}{\gamma -1},
\end{equation}
which for $z = H/2$ leads to the same result as for the VSI $\tau_c$.
An approximate growth rate for GSF at arbitrary thermal relaxation rates is then given as
\begin{equation}
    \Gamma_\mathrm{GSF} = \frac{h^2 q^2}{4} \frac{\gamma}{\gamma -1 }\frac{1}{\tau^*_\mathrm{c,GSF} + \tau^*} \Omega.
    \label{Eq:GAMMA_GSF}
\end{equation}
The local GSF growth behavior we find here is surprisingly similar to the growth rates for COS in Equation \eqref{Eq:GAMMA_COS}. 
We plot these analytic growth rates in addition to the numerical solution of the dispersion relation in Fig.\ \ref{Fig:TAUGRATE4AQ1} and find a good agreement. The deviations for GSF result from neglecting higher order terms in the dispersion relation. For the COS we find a deviation close to the midplane in the case it is radially stable $q=-0.5$, because we did not consider that the real part of the oscillations is no longer close to $\Omega$ as the direction of the oscillation turns vertical and $N_- \rightarrow 0$ for $z \rightarrow 0$.


For sufficiently long cooling times $\tau^* > 1$ growth rates for GSF and COS differ exactly by a factor of two, independent of details of the disk structure, like $p$, $h$, $q$ or $\gamma$, as can be seen by comparison of Equation \eqref{Eq:GAMMA_COS} and Equation \eqref{Eq:GAMMA_GSF}.
\response{Note that growth rates for the GSF modes for cooling times longer than the critical value scale as $q^2 h^2$ and with $1/\tau$, but are largely height independent. The GSF growth-rates for fast cooling on the other hand, are height dependent - they scale with $z/R$ - and just as for the global VSI modes they scale as $\propto q$ \citep{Manger2021}. This highlights the differences between fast and slow cooling for GSF modes despite their close relation and identical stability criterion $\kappa_z^2 = 0$.}

\subsection{The role of the radial density stratification}
\begin{figure*}
    \centering
    \fig{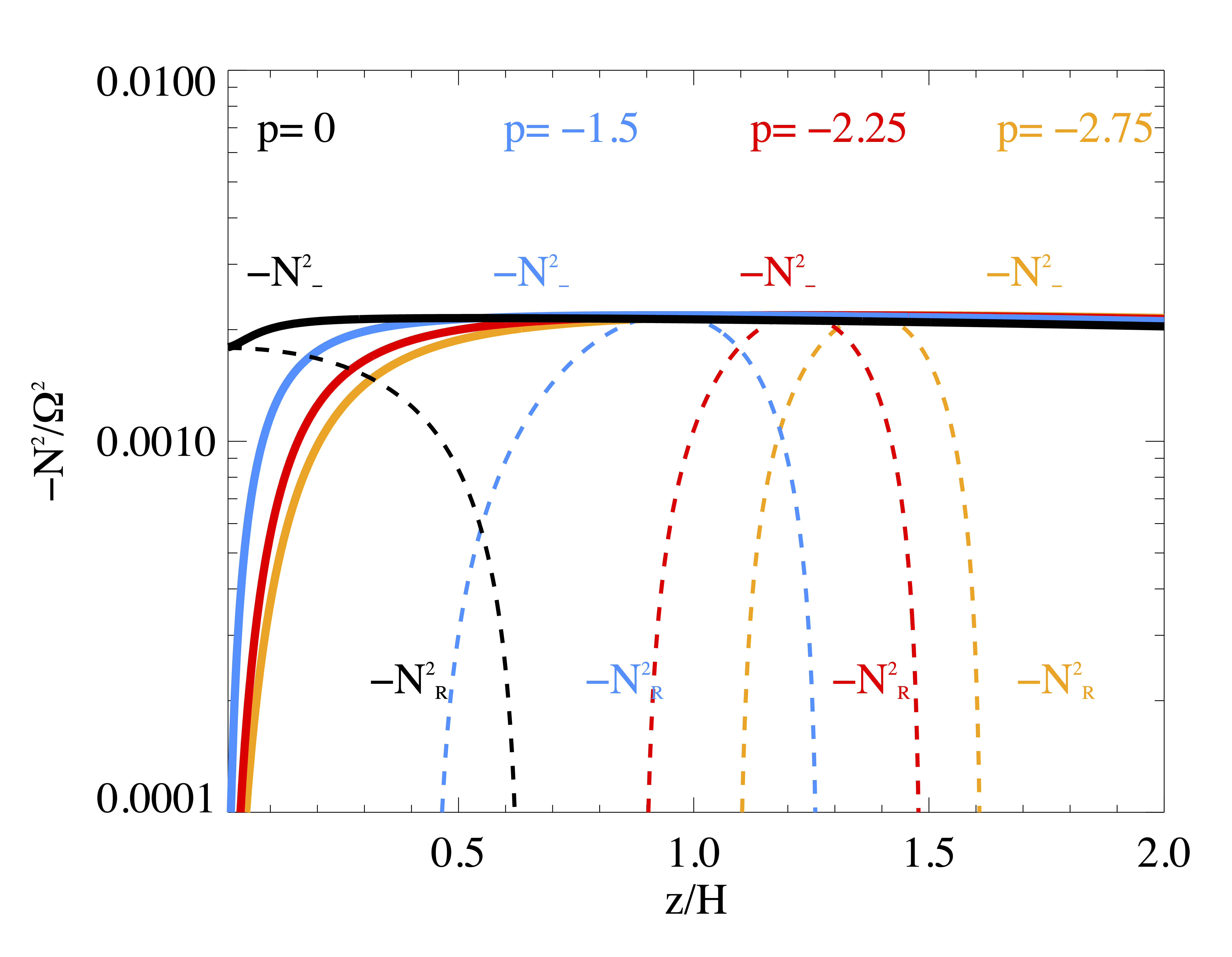}{1.0\textwidth}{}
    \caption{Most unstable directional \BVF $N^2_-$ (solid lines) as function of height $z$ above the midplane for a temperature profile of $q = -0.5$ in combination with various values for the density stratification $p$ (see colors). We also plot the corresponding radial \BVF $N_R^2$ in addition as  dashed lines. Depending on $p$ the radial $N^2_R$ is only negative (leading to radial COS) for a small vertical section of the disk atmosphere. The general COS is determined by $N^2_-$, which above $z>0.5H$ obtaines a value independent from $p$ or $z$.} 
        \label{Fig:NRP2}
\end{figure*}

\response{The growth-rates of VSI and also for GSF do not depend on the radial density stratification $p$. In contrast the growth-rates for COS in the vertically unstratified case or in the midplane for a stratified case do depend on the $p$ via the value and sign of $N^2_R$. But away from the midplane the sign and value $N^2_-$ is the important factor, and starting from a certain height, this value does not depend on the radial density stratification. In Fig.\ \ref{Fig:NRP2} we plot $N^2_-$ as function of height above the midplane for a radial temperature profile of $q=-0.5$ and four different density stratifications with $p=0$, $p=-1.5$ as well for the steeper but possibly more realistic $p=-2.25$ and $p=-2.75$ profiles.}

\response{As expected, the radial \BVF $N^2_R$ (dashed lines) varies with both pressure $p$ and height $z$. At $p = 0$, we observe radial instability $N_R^2 < 0$ at the midplane $z=0$. However, for other $p$ values, the midplane is radially stable. Nevertheless, at heights above the midplane, $N_R^2 < 0$ consistently reaches the same peak value. Remarkably, this peak value matches $N^2_-$, which remains nearly constant across a wide range of atmospheric conditions for all $p$ values. Consequently, except for the region around the midplane ($z<0.5 H$), the radial density stratification in the midplane $p$ appears insignificant in determining the occurrence and strength of COS modes.}

\begin{figure*}
    \centering
    {\fig{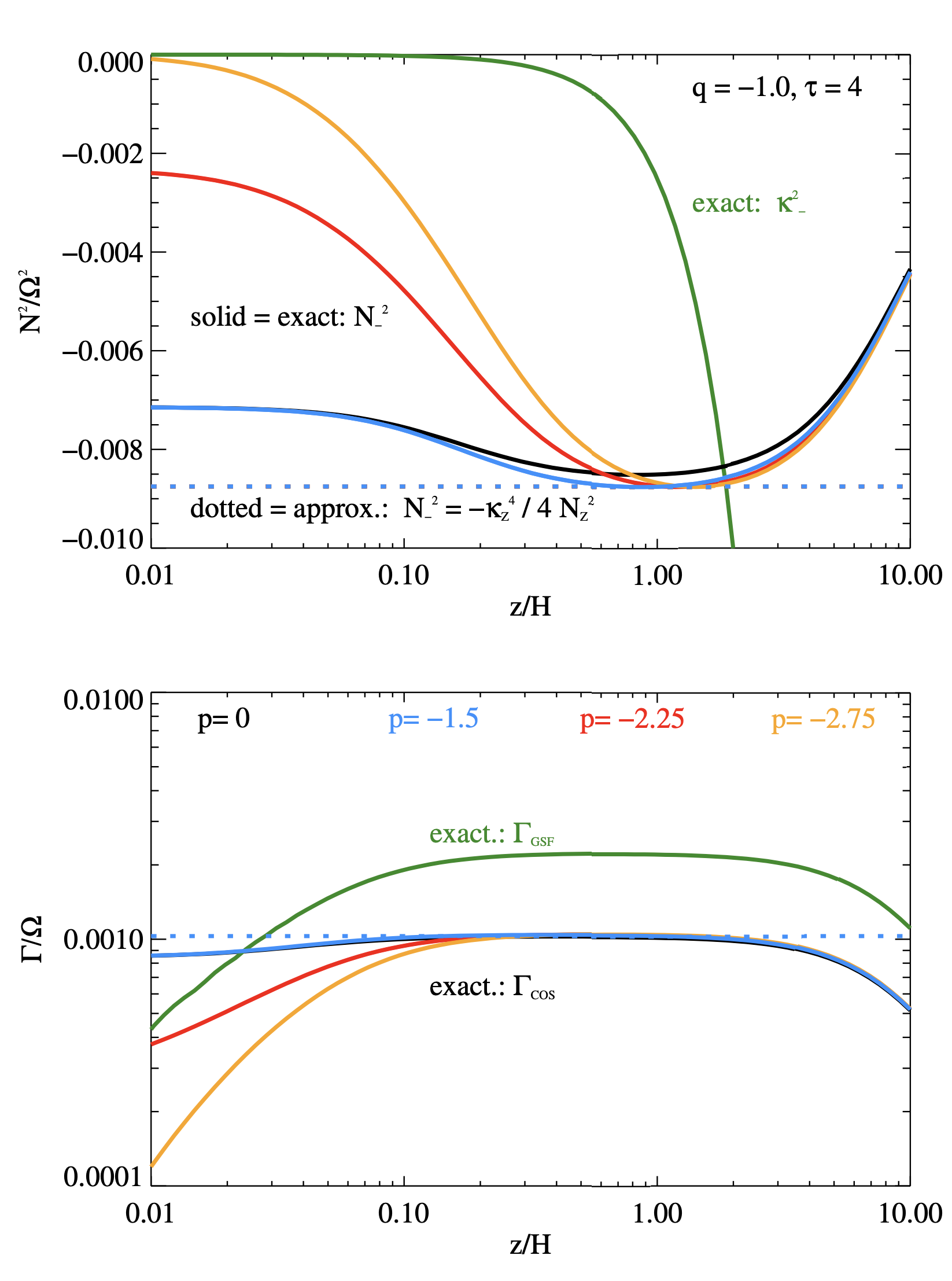}{0.5\textwidth}{(a): q = -1}\fig{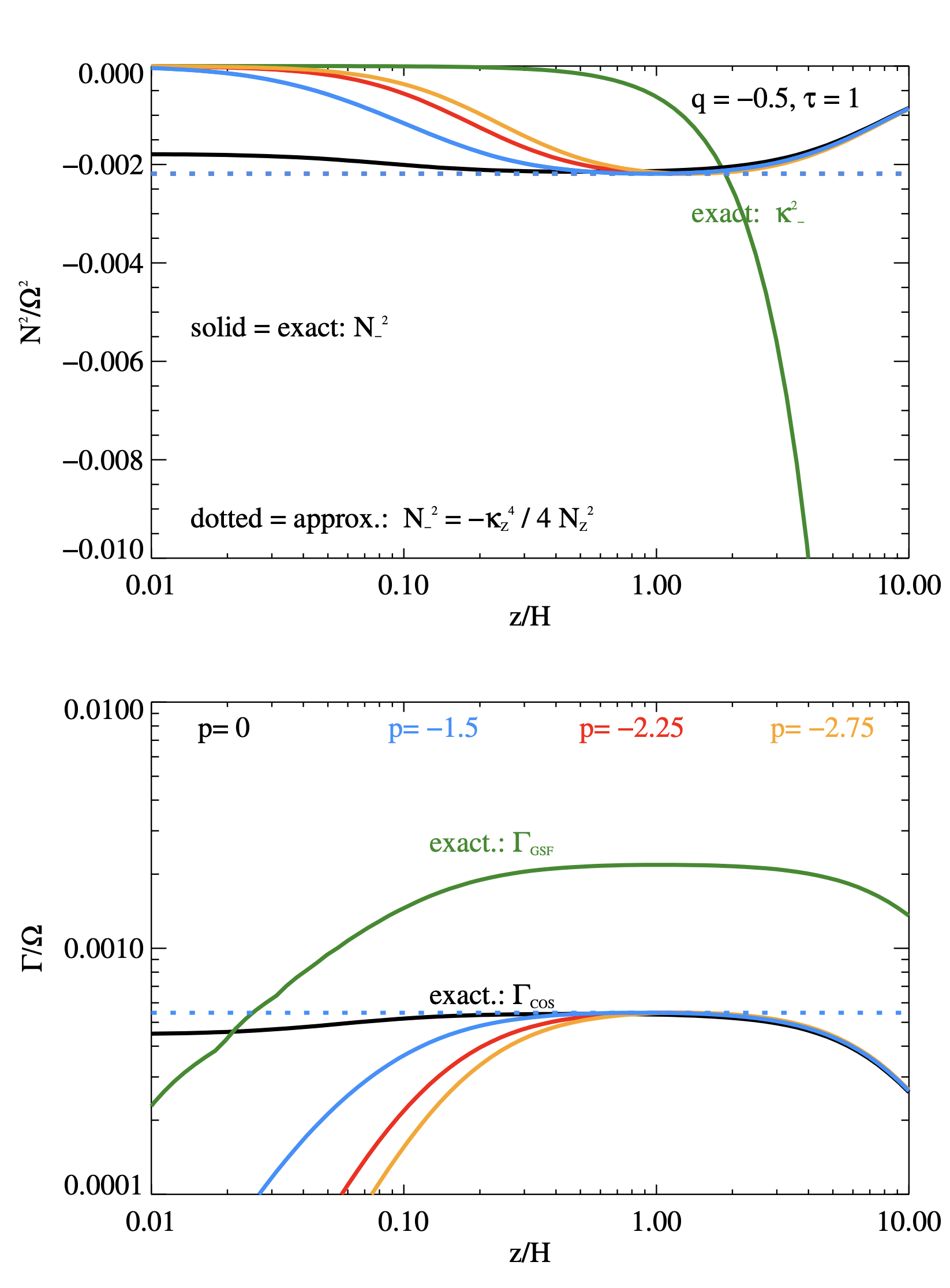}{0.5\textwidth}{(b): q = -0.5}}
    \caption{Effect of radial density stratification $q$ on the level of instability. Top row: 
    \BVFs $N^2_-$ (solid lines) as function of height $z$ above the mid-plane for a temperature profile of either $q = -0.5$ or $q = -1$ in combination with various values for the density stratification $p$ (see colors). We also plot the approximate \BVF $N_-^2$ in addition as dotted lines and the most unstable angular momentum gradient $\kappa_-^2$ as a green line. Bottom row: The resulting growth-rates for COS converge to the analytic estimate for $z > 0.1 H$ and stay at a roughly constant value up to at least $z = 4H$. For $q = -0.5$ we picked the thermal relaxation rate of $\tau^* = 1$ and for $q = -1$ a value of $\tau^* = 4$, which leads to the same GSF growth rates for both $q$ values, but twice larger COS growth rates for $q = -1$ than for $q = -0.5$.} 
        \label{Fig:N2MKPCQ}
\end{figure*}

\response{For the growth rates plotted in Fig.\ \ref{Fig:N2MKPCQ} (see also \ref{Fig:TAUGRATE4AQ1}), this implies slightly lower COS growth rates for $z< 0.2 H$. However, for larger $z$ values, the growth rates are identical.}

\begin{figure*}
    \centering
       \gridline{\fig{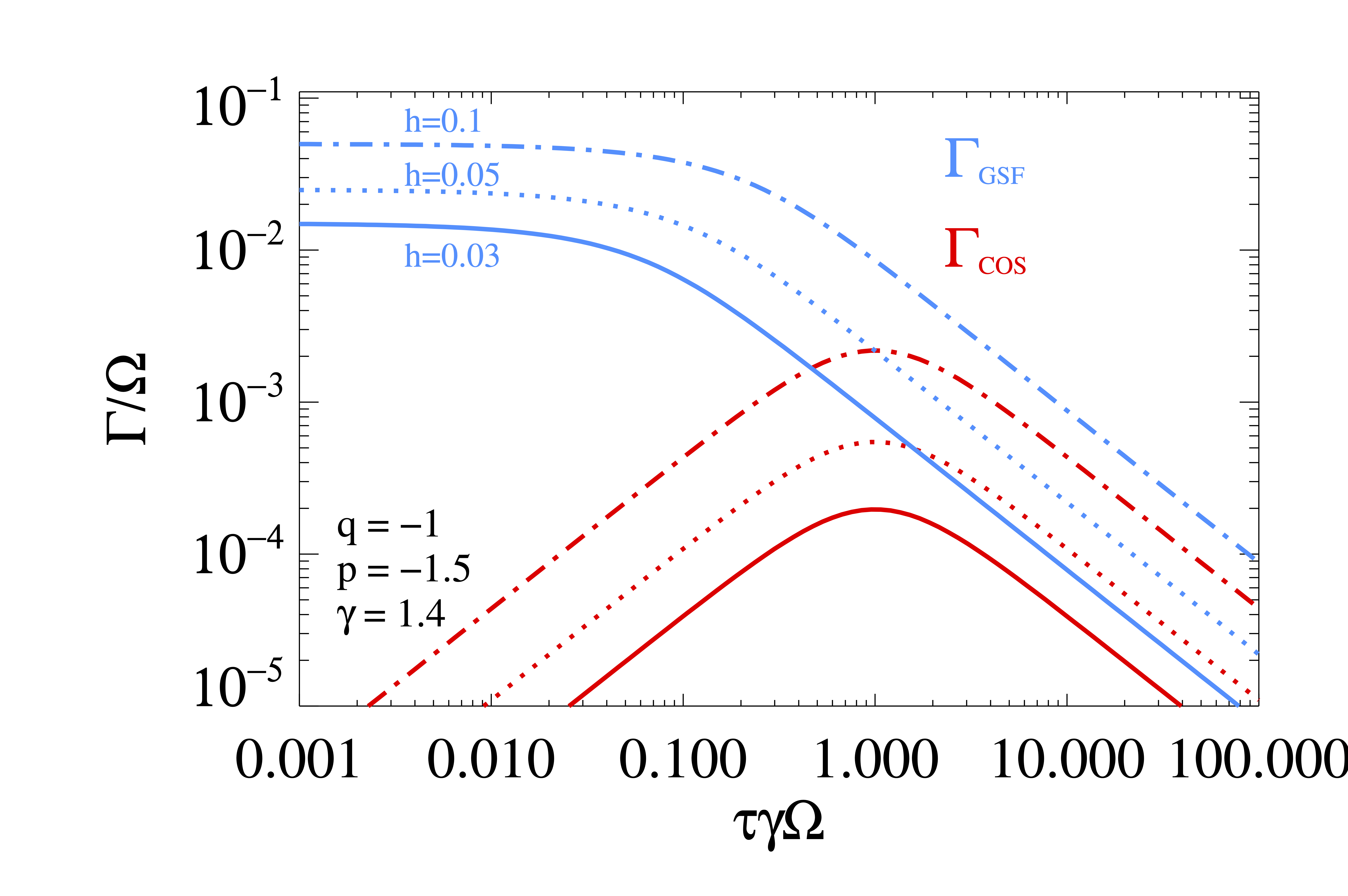}{0.5\textwidth}{(a): q = -1}\fig{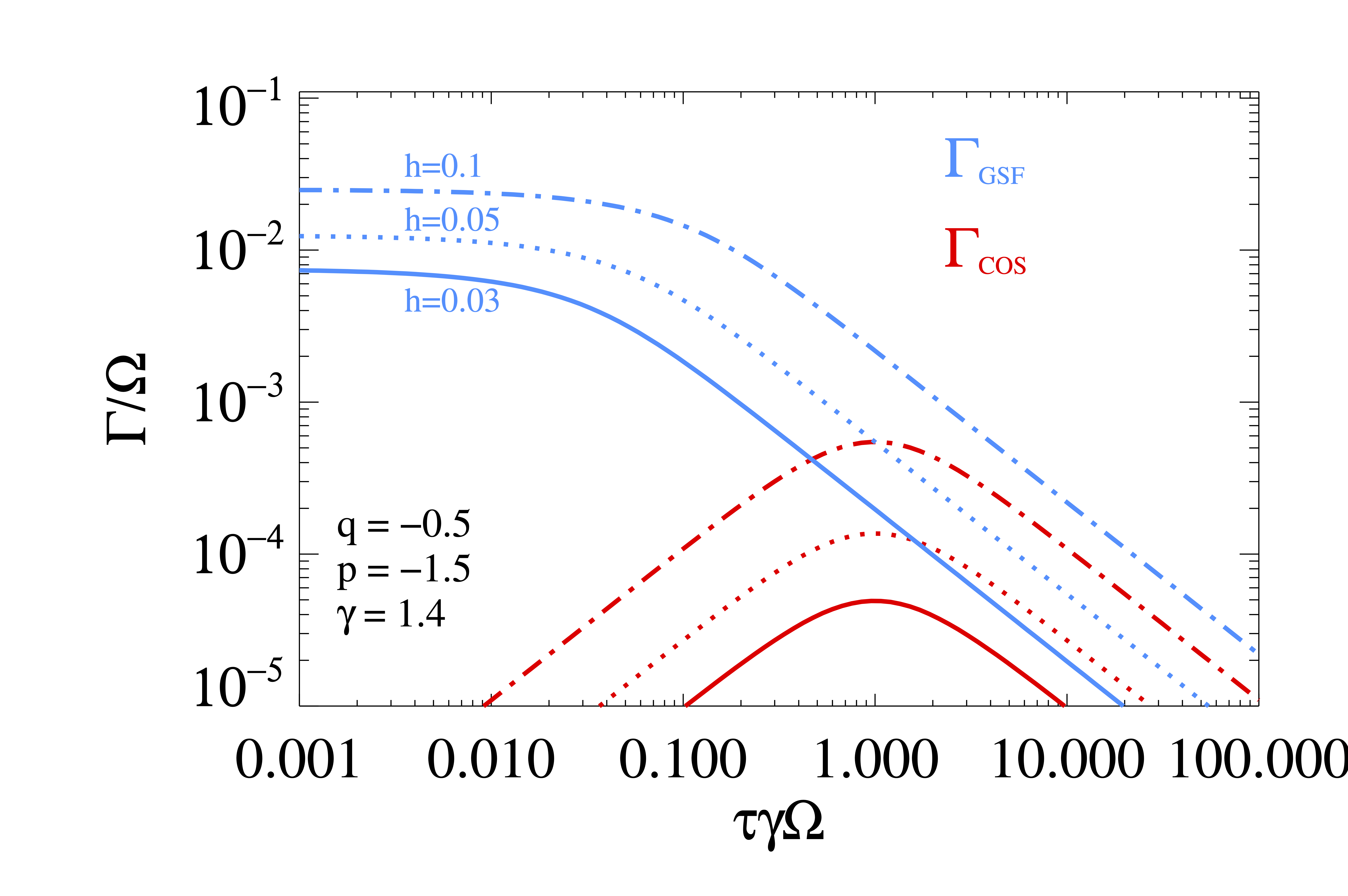}{0.5\textwidth}{(b): q = -0.5}}
    \caption{Comparison of growth rates $\Gamma$ as a function of thermal relaxation time at $z=h$ for three different values of $h = 0.1, 0.05$ and $0.03$.
    } 
        \label{Fig:TAUGRATE4AHQ1}
\end{figure*}

\response{GSF and VSI growthrates for thermal relaxation shorter than the critical timescale $\tau_c$ scale linear with pressure scale height $h$ and $q$. But for longer thermal relaxation times both VSI and GSF growth-rates scale with $h^2$ and $q^2$. In Fig.\ \ref{Fig:TAUGRATE4AHQ1} we compare growthrates for various scale height values $h=0.1$, $h=0.05$ and $h=0.03$.}

\section{Discussion}
This paper explores the local stability of stratified, differentially rotating fluids to axisymmetric perturbations in the presence of thermal relaxation mechanisms such as heat conductivity, dust-gas collisions, and radiative cooling. We present a dispersion relation already documented in the literature \citep{Goldreich1967}, which aligns with the limiting cases discussed in previous works, considering scenarios without magnetic fields and viscosity \citep{Urpin1998,Balbus2001,Menou2004}.

What's novel is the identification of a thermal overstability, alongside the well-known Goldreich-Schubert-Fricke (GSF) thermal instability \citep{Shibahashi1980}, existing in all baroclinic atmospheres, termed Convective Overstability (COS). Previous studies have primarily examined COS in vertically unstratified environments \citep{Klahr2014, Lyra2014}, specifically in barotropic atmospheres.

While the basic dispersion relation and growth rate equations were similarly derived in \citet{Urpin2003} for accretion disks, it was assumed that disks around young stars are generally convectively stable with $N^2_- > 0$. However, according to the Solberg-Høiland criteria \citep{Tassoul2000}, disks are not convectively unstable in the absence of thermal relaxation, i.e., they exhibit no dynamic instability.

The condition for thermal overstability in a baroclinic atmosphere under thermal relaxation is not determined solely by the radial and vertical buoyancy frequencies (\BVFs), i.e., $N_R^2 + N_z^2 < 0$, because $N^2_R$ and $N^2_z$ are not components of a vector \citep{Balbus1991}. Instead, it is dictated by the lowest buoyancy frequency in any perturbation direction.

\begin{figure*}
    \centering
    \fig{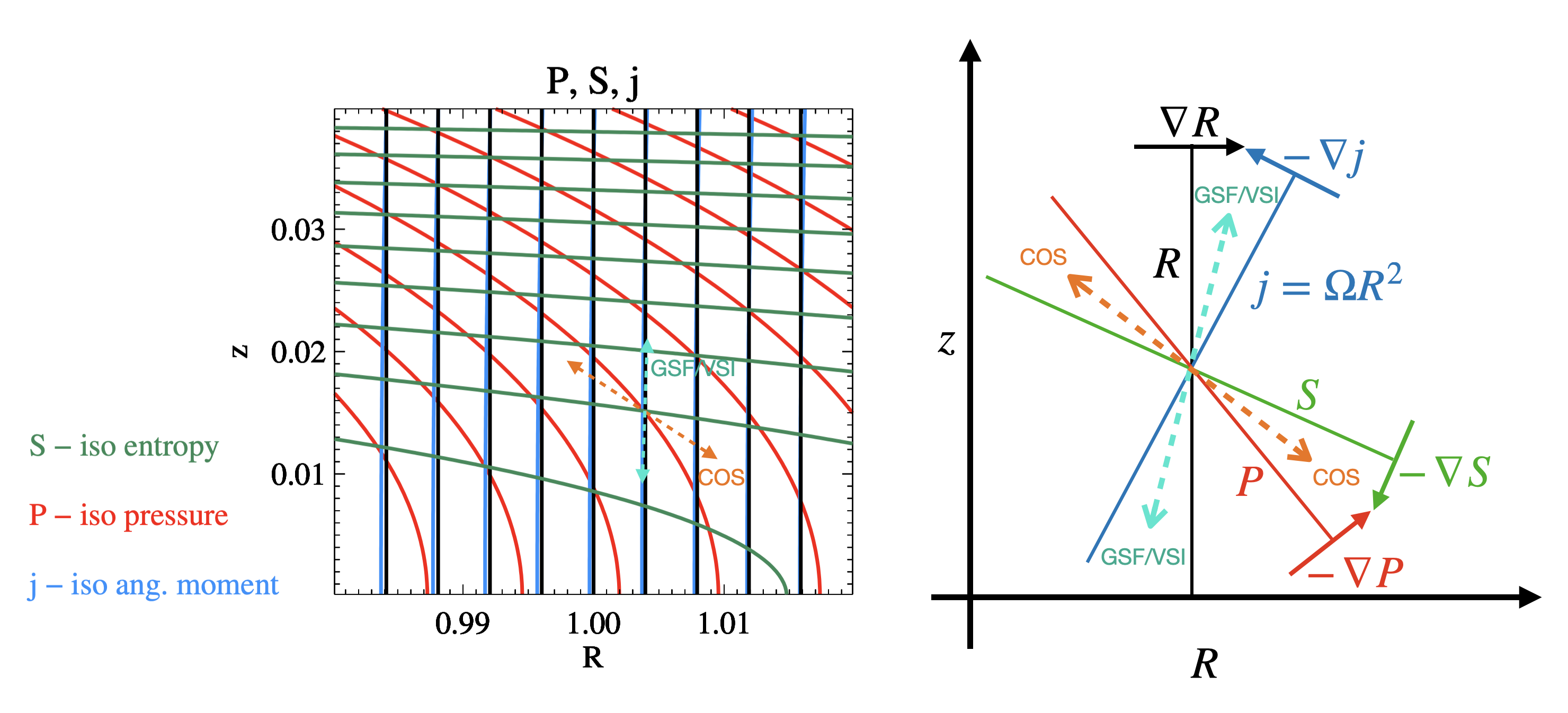}{1.0\textwidth}{}
    \caption{Gradients in a baroclinic disk atmosphere. Left plot: iso-contours of pressure (red), entropy (green) and angular momentum (blue) in a disk with $q = -0.5$,$p=-1.5$ and $h=0.1$. Additionally we plot the vertical lines of constant cylindrical radius (black) to show that the angular momentum is not constant with height. The green dashed arrows indicate the direction of unstable GSF modes. In the plotted region the entropy increases and  the pressure decrease with radius, yet, as indicated by the orange dashed arrows, there is a diagonal direction in which both decrease (or both increase), thus the necessary condition for COS is given. Right plot: To better illustrate the situation at $z=0.2 H$ we exaggerate the direction of gradients slightly and give them equal length for better illustration. A very similar figure can be found in \citet{Tassoul2000}(page 85), explaining the conditions for thermal instabilities. The black, blue, red and green lines are contours of constant radius $R$, constant specific angular momentum $j$, constant specific entropy $S$ and constant pressure $P$. We also indicate the gradients of those quantities. The direction in which specific angular momentum decreases with increasing radius is the GSF/VSI region, cyan dashed arrows, or thermally unstable situation. Thermal overstability (COS) will occur in the quasi orthogonal direction, where entropy and pressure simultaneously decrease (or increase) for motions along the orange dashed double arrow. The red/green overstable part can also be found in \citet{Kitchatinov2014} as well as in \citet{Knobloch1983}. Note that the radial component of the entropy gradient does not have to point in the same direction as the radial component of the pressure gradient, thus $N_R^2 > 0$ is not a sufficient condition for stability.} 
        \label{Fig:Grad}
\end{figure*}

Our study demonstrates that all baroclinic atmospheres exhibit a range of positive and negative values for $N^2$ depending on the direction of perturbation considered. The baroclinicity within the atmosphere induces vertical shear, expressed as the gradient of the square of the angular momentum:
\begin{equation}
    \kappa_z^2 = R \partial_z \Omega^2 = \frac{1}{\rho^2} \mathbf{\nabla} \rho \times \mathbf{\nabla} P.
\end{equation}
This term acts as the catalyst for GSF and VSI instability because it signifies a specific direction in which a centrifugal instability arises, characterized by a negative square of the oscillation frequency (refer to Fig.\ \ref{Fig:Grad}):
\begin{equation}
\kappa_-^2 = - \frac{\kappa_z^4}{4 \Omega^2}.
\end{equation}
We can demonstrate that when $\kappa_z^2 \ne 0$, it logically follows that there must a buoyantly unstable stratified direction with $N^2_- < 0$ (see Fig.\ \ref{Fig:Grad}).
For a vertically isothermal atmosphere where $N^2_z > 0$, a satisfactory approximation is provided as:
\begin{equation}
    N_-^2 = - \frac{\kappa_-^2 }{N^2_z} \Omega^2.
\end{equation}
For cooling times exceeding the critical time for VSI, $\tau > \tau_c \approx h \Omega^{-1}$, growth rates for both GSF and COS are comparable in strength, being approximately 10-100 times longer than for instantaneous cooling VSI. The peak growth for COS occurs when $\tau \gamma \Omega = 1$. As cooling times increase, GSF rates become precisely twice as fast as COS modes, highlighting the interconnected nature of instability conditions.

Thus, the stability criterion for thermal instability (GSF) and thermal overstability (COS) is $\kappa_z^2 = 0$ and likewise $q=0$ for verically isothermal disks. Additionally, for COS, stability requires $N_R^2 \geq 0$, which can be significant near the midplane, particularly when chemical gradients or dust sedimentation contribute to additional stabilization of the vertical GSF modes.

\response{For small wave-length VSI modes ($k_R H \le 10$) \citet{Lin2015} found marginal stability at thermal relaxation times longer than the critical value $\tau_c$. Only smaller modes ($k_R H \ge 30$) were still growing, which qualitatively fits our results on local GSF modes. A quantitative comparison between our analytic expressions and the numerical values in \citet{Lin2015} Fig.\ (13) could deliver interesting results.
What we find in the present paper is that not only GSF, but also COS modes can fill the potential gap between global VSI modes $\tau < 0.1 \Omega^{-1}$ and Zombie vortices $\tau > 10 \Omega^{-1}$ \citep{Barranco2018}.}
 

\response{Our numerical simulations (see paper II: \citet{Klahr2023}) show how GSF and VSI manifest themselves in the atmosphere of an accretion disk and that we can reproduce linear growth-rates down to a level of $\Gamma = 10^{-4} \Omega$ confirming analytic predictions made in this paper.}

\response{The predicted overall low growth rates will ask for long simulation times at high resolution and low dissipation, as much as for patience in performing these studies. Yet a long growth time does not necessarily predict at which level of turbulence the non-linear stage of these thermal baroclinic instabilities will saturate. Future simulations will have to clarify this. This paper along with its companion \citep{Klahr2023} therefore serves as a warning that the non-existence of turbulence for longer cooling times as reported in \citet{Manger2021} does not mean that there is no instability, but rather indicates that resolution was too low and the numerical scheme too dissipative.}

\response{In light of our progress in modeling the detailed thermal balance of disks regulated by opacities and dust gas collisions for disks around young stars \citep{MelonFuksman2024a,MelonFuksman2024b,Muley2023,Pfeil2023} it is important to have detailed predictions for growth rates to test the complex numerical schemes employed to study hydrodynamical instabilities in planet-forming disks.}

\response{As a final remark, we want to stress that the importance of these thermal baroclinic instabilities is probably not to drive the global viscous evolution of the disk but rather to set the stage for planet formation \citep{Klahr2018}. Low-level turbulence is preferable to both no turbulence and strong turbulence likewise. Excessive turbulence drives fragmentation and opposes sedimentation, thus hindering the achievement of the necessary dust-to-gas ratios and pebble sizes to trigger Streaming Instabilities (SI) and the gravitational collapse of pebble clouds \citep{Klahr2021,Lim2023}. In the absence of global turbulence, the necessary dust-to-gas ratios for SI might also never be reached \citep{Johansen2007}. However, low-level turbulence may (A) allow for large pebbles and (B) drive an inverse cascade leading to vortices \citep{Raettig2015,Raettig2021} and zonal flows \citep{Johansen2009,Dittrich2013}, which act as pebble traps to facilitate planetesimal formation \citep{Carrera2022}.}

\section*{Acknowledgments}
The author wishes to thank Hans Baehr, David Melon Fuksman, Natascha Manger, Orkan Umurhan, and Wladimir Lyra for providing feedback on early versions of this manuscript. The author is supported by the German Science Foundation (DFG) under the priority program SPP 1992: “Exoplanet Diversity” under
contract KL 1469/16-1/2. Simulations leading to the presented results were performed on the ISAAC and VERA clusters of the MPIA and the COBRA, HYDRA and RAVEN clusters of the Max-Planck-Society, both hosted at the Max-Planck Computing and Data Facility in Garching (Germany). We also acknowledge support from the DFG via the Heidelberg Cluster of Excellence STRUCTURES in the framework of Germany's Excellence Strategy (grant EXC-2181/1 - 390900948).


\begin{thebibliography}{}
\expandafter\ifx\csname natexlab\endcsname\relax\def\natexlab#1{#1}\fi
\providecommand{\url}[1]{\href{#1}{#1}}
\providecommand{\dodoi}[1]{doi:~\href{http://doi.org/#1}{\nolinkurl{#1}}}
\providecommand{\doeprint}[1]{\href{http://ascl.net/#1}{\nolinkurl{http://ascl.net/#1}}}
\providecommand{\doarXiv}[1]{\href{https://arxiv.org/abs/#1}{\nolinkurl{https://arxiv.org/abs/#1}}}

\bibitem[{{Balbus}(2001)}]{Balbus2001}
{Balbus}, S.~A. 2001, \apj, 562, 909, \dodoi{10.1086/323875}

\bibitem[{{Balbus} \& {Hawley}(1991)}]{Balbus1991}
{Balbus}, S.~A., \& {Hawley}, J.~F. 1991, The Astrophysical Journal, 376, 214,
  \dodoi{10.1086/170270}

\bibitem[{{Barker} \& {Latter}(2015)}]{Barker2015}
{Barker}, A.~J., \& {Latter}, H.~N. 2015, \mnras, 450, 21,
  \dodoi{10.1093/mnras/stv640}

\bibitem[{{Barranco} {et~al.}(2018){Barranco}, {Pei}, \&
  {Marcus}}]{Barranco2018}
{Barranco}, J.~A., {Pei}, S., \& {Marcus}, P.~S. 2018, \apj, 869, 127,
  \dodoi{10.3847/1538-4357/aaec80}

\bibitem[{{Caleo} {et~al.}(2016){Caleo}, {Balbus}, \& {Tognelli}}]{Caleo2016}
{Caleo}, A., {Balbus}, S.~A., \& {Tognelli}, E. 2016, \mnras, 460, 338,
  \dodoi{10.1093/mnras/stw1002}

\bibitem[{{Carrera} {et~al.}(2022){Carrera}, {Thomas}, {Simon}, {Small},
  {Kretke}, \& {Klahr}}]{Carrera2022}
{Carrera}, D., {Thomas}, A.~J., {Simon}, J.~B., {et~al.} 2022, \apj, 927, 52,
  \dodoi{10.3847/1538-4357/ac4d28}

\bibitem[{{Charney}(1947)}]{Charney1947}
{Charney}, J.~G. 1947, Journal of Atmospheric Sciences, 4, 136,
  \dodoi{10.1175/1520-0469(1947)004<0136:TDOLWI>2.0.CO;2}

\bibitem[{{Cui} \& {Latter}(2022)}]{Cui2022}
{Cui}, C., \& {Latter}, H. 2022, arXiv e-prints, arXiv:2201.09038.
\newblock \doarXiv{2201.09038}

\bibitem[{{Dittrich} {et~al.}(2013){Dittrich}, {Klahr}, \&
  {Johansen}}]{Dittrich2013}
{Dittrich}, K., {Klahr}, H., \& {Johansen}, A. 2013, \apj, 763, 117,
  \dodoi{10.1088/0004-637X/763/2/117}

\bibitem[{{Eady}(1949)}]{Eady1949}
{Eady}, E.~T. 1949, Tellus, 1, 33, \dodoi{10.3402/tellusa.v1i3.8507}

\bibitem[{{Fricke}(1968)}]{Fricke1968}
{Fricke}, K. 1968, Zeitschrift f{\"u}r Astrophysik, 68, 317

\bibitem[{{Garaud}(2021)}]{Garaud2021}
{Garaud}, P. 2021, Physical Review Fluids, 6, 030501,
  \dodoi{10.1103/PhysRevFluids.6.030501}

\bibitem[{{Goldreich} \& {Lynden-Bell}(1965)}]{1965MNRAS.130..125G}
{Goldreich}, P., \& {Lynden-Bell}, D. 1965, \mnras, 130, 125,
  \dodoi{10.1093/mnras/130.2.125}

\bibitem[{{Goldreich} \& {Schubert}(1967)}]{Goldreich1967}
{Goldreich}, P., \& {Schubert}, G. 1967, The Astrophysical Journal, 150, 571,
  \dodoi{10.1086/149360}

\bibitem[{{Hawley} \& {Balbus}(1991)}]{1991ApJ...376..223H}
{Hawley}, J.~F., \& {Balbus}, S.~A. 1991, \apj, 376, 223,
  \dodoi{10.1086/170271}

\bibitem[{Holton \& Hakim(2012)}]{Holton2012-kg}
Holton, J.~R., \& Hakim, G.~J. 2012, An introduction to dynamic meteorology,
  5th edn., International Geophysics (Academic Press)

\bibitem[{{Howard}(1961)}]{Howard1961}
{Howard}, L.~N. 1961, Journal of Fluid Mechanics, 10, 509,
  \dodoi{10.1017/S0022112061000317}

\bibitem[{{Johansen} {et~al.}(2007){Johansen}, {Oishi}, {Mac Low}, {Klahr},
  {Henning}, \& {Youdin}}]{Johansen2007}
{Johansen}, A., {Oishi}, J.~S., {Mac Low}, M.-M., {et~al.} 2007, Nature, 448,
  1022, \dodoi{10.1038/nature06086}

\bibitem[{{Johansen} {et~al.}(2009){Johansen}, {Youdin}, \&
  {Klahr}}]{Johansen2009}
{Johansen}, A., {Youdin}, A., \& {Klahr}, H. 2009, \apj, 697, 1269,
  \dodoi{10.1088/0004-637X/697/2/1269}

\bibitem[{{Kato}(1966)}]{Kato1966}
{Kato}, S. 1966, \pasj, 18, 374

\bibitem[{{Kitchatinov}(2014)}]{Kitchatinov2014}
{Kitchatinov}, L.~L. 2014, \apj, 784, 81, \dodoi{10.1088/0004-637X/784/1/81}

\bibitem[{Klahr(2004)}]{Klahr2004}
Klahr, H. 2004, The Astrophysical Journal, 606, 1070, \dodoi{10.1086/383119}

\bibitem[{{Klahr} {et~al.}(2023){Klahr}, {Baehr}, \& {Melon
  Fuksman}}]{Klahr2023}
{Klahr}, H., {Baehr}, H., \& {Melon Fuksman}, J.~D. 2023, arXiv e-prints,
  arXiv:2305.08165, \dodoi{10.48550/arXiv.2305.08165}

\bibitem[{{Klahr} \& {Hubbard}(2014)}]{Klahr2014}
{Klahr}, H., \& {Hubbard}, A. 2014, The Astrophysical Journal, 788, 21,
  \dodoi{10.1088/0004-637X/788/1/21}

\bibitem[{Klahr {et~al.}(2018)Klahr, Pfeil, \& Schreiber}]{Klahr2018}
Klahr, H., Pfeil, T., \& Schreiber, A. 2018, in Handbook of Exoplanets (Cham:
  Springer International Publishing), 1--36.
\newblock
  \url{https://link.springer.com/referenceworkentry/10.1007%2F978-3-319-55333-7_138#enumeration}

\bibitem[{{Klahr} \& {Schreiber}(2021)}]{Klahr2021}
{Klahr}, H., \& {Schreiber}, A. 2021, \apj, 911, 9,
  \dodoi{10.3847/1538-4357/abca9b}

\bibitem[{{Klahr} \& {Bodenheimer}(2003)}]{Klahr2003}
{Klahr}, H.~H., \& {Bodenheimer}, P. 2003, The Astrophysical Journal, 582, 869,
  \dodoi{10.1086/344743}

\bibitem[{{Knobloch} \& {Spruit}(1983)}]{Knobloch1983}
{Knobloch}, E., \& {Spruit}, H.~C. 1983, \aap, 125, 59

\bibitem[{{Knobloch} \& {Spruit}(1986)}]{Knobloch1986}
---. 1986, \aap, 166, 359

\bibitem[{{Latter} \& {Kunz}(2022)}]{Latter2022}
{Latter}, H.~N., \& {Kunz}, M.~W. 2022, \mnras, \dodoi{10.1093/mnras/stac107}

\bibitem[{{Latter} \& {Papaloizou}(2017)}]{Latter2017}
{Latter}, H.~N., \& {Papaloizou}, J. 2017, \mnras, 472, 1432,
  \dodoi{10.1093/mnras/stx2038}

\bibitem[{{Le Bars} \& {Le Gal}(2007)}]{LeBars2007}
{Le Bars}, M., \& {Le Gal}, P. 2007, \prl, 99, 064502,
  \dodoi{10.1103/PhysRevLett.99.064502}

\bibitem[{{Lesur} \& {Papaloizou}(2009)}]{Lesur2009}
{Lesur}, G., \& {Papaloizou}, J.~C.~B. 2009, Astronomy and Astrophysics, 498,
  1, \dodoi{10.1051/0004-6361/200811577}

\bibitem[{{Lesur} {et~al.}(2023){Lesur}, {Flock}, {Ercolano}, {Lin}, {Yang},
  {Barranco}, {Benitez-Llambay}, {Goodman}, {Johansen}, {Klahr}, {Laibe},
  {Lyra}, {Marcus}, {Nelson}, {Squire}, {Simon}, {Turner}, {Umurhan}, \&
  {Youdin}}]{Lesur2023}
{Lesur}, G., {Flock}, M., {Ercolano}, B., {et~al.} 2023, in Astronomical
  Society of the Pacific Conference Series, Vol. 534, Protostars and Planets
  VII, ed. S.~{Inutsuka}, Y.~{Aikawa}, T.~{Muto}, K.~{Tomida}, \& M.~{Tamura},
  465

\bibitem[{{Lim} {et~al.}(2023){Lim}, {Simon}, {Li}, {Armitage}, {Carrera},
  {Lyra}, {Rea}, {Yang}, \& {Youdin}}]{Lim2023}
{Lim}, J., {Simon}, J.~B., {Li}, R., {et~al.} 2023, arXiv e-prints,
  arXiv:2312.12508, \dodoi{10.48550/arXiv.2312.12508}

\bibitem[{{Lin}(2019)}]{Lin2019}
{Lin}, M.-K. 2019, \mnras, 485, 5221, \dodoi{10.1093/mnras/stz701}

\bibitem[{{Lin} \& {Youdin}(2015)}]{Lin2015}
{Lin}, M.-K., \& {Youdin}, A.~N. 2015, The Astrophysical Journal, 811, 17,
  \dodoi{10.1088/0004-637X/811/1/17}

\bibitem[{{Lovelace} {et~al.}(1999){Lovelace}, {Li}, {Colgate}, \&
  {Nelson}}]{Lovelace1999}
{Lovelace}, R.~V.~E., {Li}, H., {Colgate}, S.~A., \& {Nelson}, A.~F. 1999,
  \apj, 513, 805, \dodoi{10.1086/306900}

\bibitem[{Lyra(2014)}]{Lyra2014}
Lyra, W. 2014, The Astrophysical Journal, 789, 77,
  \dodoi{10.1088/0004-637x/789/1/77}

\bibitem[{{Lyra} \& {Klahr}(2011)}]{Lyra2011}
{Lyra}, W., \& {Klahr}, H. 2011, \aap, 527, A138,
  \dodoi{10.1051/0004-6361/201015568}

\bibitem[{{Malygin} {et~al.}(2017){Malygin}, {Klahr}, {Semenov}, {Henning}, \&
  {Dullemond}}]{Malygin2017}
{Malygin}, M.~G., {Klahr}, H., {Semenov}, D., {Henning}, T., \& {Dullemond},
  C.~P. 2017, Astronomy \& Astrophysics, 605, A30,
  \dodoi{10.1051/0004-6361/201629933}

\bibitem[{{Manger} {et~al.}(2021){Manger}, {Pfeil}, \& {Klahr}}]{Manger2021}
{Manger}, N., {Pfeil}, T., \& {Klahr}, H. 2021, \mnras, 508, 5402,
  \dodoi{10.1093/mnras/stab2599}

\bibitem[{{Marcus} {et~al.}(2016){Marcus}, {Pei}, {Jiang}, \&
  {Barranco}}]{Marcus2016}
{Marcus}, P.~S., {Pei}, S., {Jiang}, C.-H., \& {Barranco}, J.~A. 2016, \apj,
  833, 148, \dodoi{10.3847/1538-4357/833/2/148}

\bibitem[{{Marcus} {et~al.}(2015){Marcus}, {Pei}, {Jiang}, {Barranco},
  {Hassanzadeh}, \& {Lecoanet}}]{Marcus2015}
{Marcus}, P.~S., {Pei}, S., {Jiang}, C.-H., {et~al.} 2015, \apj, 808, 87,
  \dodoi{10.1088/0004-637X/808/1/87}

\bibitem[{{McIntyre}(1970)}]{McIntyre1970}
{McIntyre}, M. 1970, Geophysical and Astrophysical Fluid Dynamics, 1, 19,
  \dodoi{10.1080/03091927009365767}

\bibitem[{{Melon Fuksman} {et~al.}(2024{\natexlab{a}}){Melon Fuksman}, {Flock},
  \& {Klahr}}]{MelonFuksman2024a}
{Melon Fuksman}, J.~D., {Flock}, M., \& {Klahr}, H. 2024{\natexlab{a}}, \aap,
  682, A139, \dodoi{10.1051/0004-6361/202346554}

\bibitem[{{Melon Fuksman} {et~al.}(2024{\natexlab{b}}){Melon Fuksman}, {Flock},
  \& {Klahr}}]{MelonFuksman2024b}
---. 2024{\natexlab{b}}, \aap, 682, A140, \dodoi{10.1051/0004-6361/202346555}

\bibitem[{{Menou} {et~al.}(2004){Menou}, {Balbus}, \& {Spruit}}]{Menou2004}
{Menou}, K., {Balbus}, S.~A., \& {Spruit}, H.~C. 2004, \apj, 607, 564,
  \dodoi{10.1086/383463}

\bibitem[{{Miles}(1961)}]{Miles1961}
{Miles}, J.~W. 1961, Journal of Fluid Mechanics, 10, 496,
  \dodoi{10.1017/S0022112061000305}

\bibitem[{{Mohandas} \& {Pessah}(2015)}]{MohandasPessah2015}
{Mohandas}, G., \& {Pessah}, M.~E. 2015, arXiv e-prints, arXiv:1510.02729.
\newblock \doarXiv{1510.02729}

\bibitem[{{Muley} {et~al.}(2023){Muley}, {Melon Fuksman}, \&
  {Klahr}}]{Muley2023}
{Muley}, D., {Melon Fuksman}, J.~D., \& {Klahr}, H. 2023, \aap, 678, A162,
  \dodoi{10.1051/0004-6361/202347101}

\bibitem[{{Nelson} {et~al.}(2013){Nelson}, {Gressel}, \&
  {Umurhan}}]{Nelson2013}
{Nelson}, R.~P., {Gressel}, O., \& {Umurhan}, O.~M. 2013, Monthly Notices of
  the Royal Astronomical Society, 435, 2610, \dodoi{10.1093/mnras/stt1475}

\bibitem[{{Nelson} {et~al.}(2016){Nelson}, {Gressel}, \&
  {Umurhan}}]{Nelson+2016}
---. 2016, \mnras, 456, 239, \dodoi{10.1093/mnras/stv2440}

\bibitem[{{Parker} {et~al.}(2021){Parker}, {Caulfield}, \&
  {Kerswell}}]{Parker2021}
{Parker}, J.~P., {Caulfield}, C.~P., \& {Kerswell}, R.~R. 2021, Journal of
  Fluid Mechanics, 915, A37, \dodoi{10.1017/jfm.2021.125}

\bibitem[{Petersen {et~al.}(2007{\natexlab{a}})Petersen, Julien, \&
  Stewart}]{Petersen2007a}
Petersen, M.~R., Julien, K., \& Stewart, G.~R. 2007{\natexlab{a}}, The
  Astrophysical Journal, 658, 1236, \dodoi{10.1086/511513}

\bibitem[{Petersen {et~al.}(2007{\natexlab{b}})Petersen, Stewart, \&
  Julien}]{Petersen2007b}
Petersen, M.~R., Stewart, G.~R., \& Julien, K. 2007{\natexlab{b}}, The
  Astrophysical Journal, 658, 1252, \dodoi{10.1086/511523}

\bibitem[{{Pfeil} {et~al.}(2023){Pfeil}, {Birnstiel}, \& {Klahr}}]{Pfeil2023}
{Pfeil}, T., {Birnstiel}, T., \& {Klahr}, H. 2023, \apj, 959, 121,
  \dodoi{10.3847/1538-4357/ad00af}

\bibitem[{{Pfeil} \& {Klahr}(2019)}]{Pfeil2019}
{Pfeil}, T., \& {Klahr}, H. 2019, The Astrophysical Journal, 871, 150,
  \dodoi{10.3847/1538-4357/aaf962}

\bibitem[{{Raettig} {et~al.}(2015){Raettig}, {Klahr}, \& {Lyra}}]{Raettig2015}
{Raettig}, N., {Klahr}, H., \& {Lyra}, W. 2015, \apj, 804, 35,
  \dodoi{10.1088/0004-637X/804/1/35}

\bibitem[{{Raettig} {et~al.}(2013){Raettig}, {Lyra}, \& {Klahr}}]{Raettig2013}
{Raettig}, N., {Lyra}, W., \& {Klahr}, H. 2013, \apj, 765, 115,
  \dodoi{10.1088/0004-637X/765/2/115}

\bibitem[{{Raettig} {et~al.}(2021){Raettig}, {Lyra}, \& {Klahr}}]{Raettig2021}
---. 2021, \apj, 913, 92, \dodoi{10.3847/1538-4357/abf739}

\bibitem[{{R{\"u}diger} {et~al.}(2002){R{\"u}diger}, {Arlt}, \&
  {Shalybkov}}]{Ruediger2002}
{R{\"u}diger}, G., {Arlt}, R., \& {Shalybkov}, D. 2002, \aap, 391, 781,
  \dodoi{10.1051/0004-6361:20020853}

\bibitem[{{Shalybkov} \& {R{\"u}diger}(2005)}]{Shalybkov2005}
{Shalybkov}, D., \& {R{\"u}diger}, G. 2005, \aap, 438, 411,
  \dodoi{10.1051/0004-6361:20042492}

\bibitem[{{Shibahashi}(1980)}]{Shibahashi1980}
{Shibahashi}, H. 1980, \pasj, 32, 341

\bibitem[{Tassoul(2000)}]{Tassoul2000}
Tassoul, J.-L. 2000, Stellar Rotation, Cambridge Astrophysics (Cambridge
  University Press), \dodoi{10.1017/CBO9780511546044}

\bibitem[{{Urpin}(2003)}]{Urpin2003}
{Urpin}, V. 2003, \aap, 404, 397, \dodoi{10.1051/0004-6361:20030513}

\bibitem[{{Urpin} \& {Brandenburg}(1998)}]{Urpin1998}
{Urpin}, V., \& {Brandenburg}, A. 1998, Monthly Notices of the Royal
  Astronomical Socienty, 294, 399, \dodoi{10.1046/j.1365-8711.1998.01118.x}

\end{thebibliography}

\bibliographystyle{aasjournal}

\appendix
\section{Disk structure and local gradients}
\label{appendix1}
We use the assumption of a vertical isothermal disk as in \citet{Nelson2013}. For a Gaussian vertical density structure with a scale height of $H$ we find:
\begin{equation}
\rho(R,z) = \rho(R,0) e^{-\frac{z^2}{2H^2}}.
\end{equation}
In addition, we find the local gradients of logarithmic density $a_z$ and pressure $b_z$ to be:
\begin{equation}
a_z = b_z = -  \frac{z}{H^2}.
\end{equation}
The radial gradients are slightly more complicated and involve the gradients of density $p$ and temperature $q$ in the midplane:
\begin{equation}
a_R = \frac{1}{R}\left(p + (3 + q) \frac{z^2}{2H^2}\right), 
\end{equation}
and
\begin{equation}
b_R = a_R + \frac{q}{R} = \frac{1}{R}\left(p + q + (3 + q) \frac{z^2}{2H^2}\right).
\end{equation}
With these values we can calculate the entropy gradients $s_R = b_R - \gamma a_R$ and $s_z = b_z - \gamma a_z$ and likewise the vertical component of the angular momentum gradient vector $\kappa_z^2$ for arbitrary locations in the disk atmosphere.

\section{Growth rates for COS}
\label{appendixCOS}
Here we briefly explain how we obtained the compact expression for the COS growthrates as given in Eq.\ \eqref{Eq:GAMMA_COS}. We start from the full dispersion relation:
\begin{equation}
%
\omega^3 + \frac{i}{\gamma \tau} \omega^2  - \omega \left[N^2_\mathbf{k} +  \kappa_\mathbf{k}^2\right]  - \frac{i}{\gamma \tau} \kappa_\mathbf{k}^2 = 0
\label{eq:ReUnstDisp0}
\end{equation}
Now we can use our knowledge about the nature of the overstable modes. They possess an oscillation frequency $\omega_0$ which is the real part of the complex frequency and complex part, indicating growth od decay rate: $\omega := \omega_0 + i \Gamma_\mathrm{COS}$, with $\Gamma_\mathrm{COS}$ the growthrates of the overstable convective oscillations. Inserting this in the dispersion relation leads to:
\begin{equation}
\omega_0^3 + 3 i \omega_0^2 \Gamma_\mathrm{COS} - 3 \omega_0  \Gamma_\mathrm{COS}^2 - i \Gamma_\mathrm{COS}^3 + \frac{i \omega_0^2}{\gamma \tau} - \frac{2 \omega_0 \Gamma_\mathrm{COS}}{\gamma \tau} - i \frac{\Gamma_\mathrm{COS}^2}{\gamma \tau} - \omega_0 \left(\kappa_k^2 + N_k^2 \right) -i \Gamma_\mathrm{COS} \left(\kappa_k^2 + N_k^2 \right) - i \frac{\kappa_k^2}{\gamma \tau} = 0,
\end{equation}
which we can separate into a real and imaginary part which independently have to be zero. The real part gives:
\begin{equation}
\omega_0^2  = 3 \Gamma_\mathrm{COS}^2  + \frac{2 \Gamma_\mathrm{COS}}{\gamma \tau}  + \left(\kappa_k^2 + N_k^2 \right)
\label{Eq:A18}
\end{equation}
and the imaginary:
\begin{equation}
3 \omega_0^2 \Gamma_\mathrm{COS} - \Gamma_\mathrm{COS}^3 + \frac{ \omega_0^2}{\gamma \tau}  -  \frac{\Gamma_\mathrm{COS}^2}{\gamma \tau}  -\Gamma_\mathrm{COS} \left(\kappa_k^2 + N_k^2 \right) -  \frac{\kappa_k^2}{\gamma \tau} = 0.
\label{Eq:B8}
\end{equation}
We insert $\omega_0^2$ from Eq. \eqref{Eq:A18}  into Eq. \eqref{Eq:B8}:
\begin{equation}
3 \left(3 \Gamma_\mathrm{COS}^2  + \frac{2 \Gamma_\mathrm{COS}}{\gamma \tau}  + \left(\kappa_k^2 + N_k^2 \right)\right) \Gamma_\mathrm{COS} - \Gamma_\mathrm{COS}^3 + \frac{3 \Gamma_\mathrm{COS}^2  + \frac{2 \Gamma_\mathrm{COS}}{\gamma \tau}  + \left(\kappa_k^2 + N_k^2 \right)}{\gamma \tau}  -  \frac{\Gamma_\mathrm{COS}^2}{\gamma \tau}  -\Gamma_\mathrm{COS} \left(\kappa_k^2 + N_k^2 \right) -  \frac{\kappa_k^2}{\gamma \tau} = 0
\end{equation}
As growthrates are small, we can drop the squared and cubed terms in $\Gamma_\mathrm{COS}$, thus:
\begin{equation}
3 \left(\kappa_k^2 + N_k^2 \right) \Gamma_\mathrm{COS}  + \frac{\frac{2 \Gamma_\mathrm{COS}}{\gamma \tau}  + \left(\kappa_k^2 + N_k^2 \right)}{\gamma \tau}   -\Gamma_\mathrm{COS} \left(\kappa_k^2 + N_k^2 \right) -  \frac{\kappa_k^2}{\gamma \tau} = 0
\end{equation}
We then rearrange the terms and multiply all terms with $\gamma^2\tau^2$:
\begin{equation}
2 \gamma^2 \tau^2 \left(\kappa_k^2 + N_k^2 \right) \Gamma_\mathrm{COS}  + 2 \Gamma_\mathrm{COS}  + \gamma \tau N_k^2 = 0,
\end{equation}
and finally find:
\begin{equation}
\Gamma_\mathrm{COS} = \frac{1}{2} \frac{- N_k^2 \gamma \tau}{1 + \gamma^2 \tau^2 \left(\kappa_k^2 + N_k^2 \right)}
\end{equation}
Inserting this result into Eq.\ \eqref{Eq:A18}, but ignoring $\Gamma^2_\mathrm{COS}$ we find the oscillation frequency to be 
\begin{equation}
\omega_0^2 = \kappa_k^2 + N^2_k \left( 1 -  \frac{1}{1 + \gamma^2 \tau^2 \left(\kappa_k^2 + N_k^2 \right)}\right).
\end{equation}
From analyzing the dispersion relation  for $\tau = 0$ we knew already that the oscillation frequency in this limit is the epicyclic frequency $\omega_0^2 = \kappa_k^2$ and in the absence of cooling we find $\omega_0^2 = \kappa_k^2 + N^2_k$. For most of the atmosphere we find $|\kappa_k^2| \gg |N^2_k|$ and we see no dependence of $\omega_0$ on $\tau$ in Fig.\ \ref{Fig:TAUGRATE4Q1}. 

\section{GSF Growth rates}
\label{AppendixGSF}
The local GSF modes are also part of the general dispersion relation,
%
but for the GSF we can assume that $\omega$ has no real part $\omega_0$. GSF is an instability, not an over-stability, thus we set $\omega = i \Gamma$ and find a dispersion relation without relevant complex solutions:
\begin{equation}
\Gamma^3  + \frac{1}{\gamma \tau} \Gamma^2  + \Gamma \left[ N_\mathbf{k}^2 + \kappa_\mathbf{k}^2\right]+ \frac{1}{\gamma \tau} \kappa_\mathbf{k}^2 = 0.
\label{eq:FullDisp2}
\end{equation}
In case of instant cooling $\tau = 0$ we get 
\begin{equation}
\Gamma^2 = - \kappa_\mathbf{k}^2 = - \frac{k_z^2}{k^2} \left(1 - \frac{k_R}{k_z} \frac{\kappa_z^2}{\Omega^2}\right) \Omega^2,
\label{eq:FullDisp3}
\end{equation}
which has a maximum in growthrate at $\partial_{k_z} \Gamma^2_0 = 0$. With $k_R \gg k_z$ it follows $(\frac{k_z}{k_R})_\mathrm{max} = \frac{\kappa_z^2}{2\Omega^2}$ and  we find:
\begin{equation}
\Gamma_\mathrm{GSF}(\tau = 0) = \frac{1}{2} \frac{|\kappa_z^2|}{\Omega^2} \Omega.
\label{eq:FullDisp4}
\end{equation}
Note that the sign of $\kappa_z^2$ does not matter, it only changes the direction of the most unstable wave vector $k_z \rightarrow -k_z$.
We can also solve Eq.\ \eqref{eq:FullDisp2} for $\tau > 0$ when ignoring the quadratic and cubic terms:
\begin{equation}
\Gamma_\mathrm{GSF}(\tau > 0) \left[N_k^2 +\kappa_\mathbf{k}^2 \right] + \frac{1}{\gamma \tau} \kappa_\mathbf{k}^2 = 0,
\label{eq:FullDisp5}
\end{equation}
which leads to:
\begin{equation}
\Gamma_\mathrm{GSF}(\tau > 0) = \frac{\kappa_\mathbf{k}^2}{\gamma \tau \left(N_z^2 - \kappa_\mathbf{k}^2\right)}.
\label{eq:FullDisp6}
\end{equation}
In the main body of the paper we show how to combine the long and short cooling time regimes.

\section{A proof}
\label{appendix2}
We want to test the general validity of the assumption:
\begin{eqnarray}
 N^2_- = \frac{\kappa_-^2}{N^2_+}\Omega^2= - \frac{1}{4} \frac{\kappa_z^4}{N^2_+},
 \label{eq:972}
\end{eqnarray}
as it would clearly indicate that for any vertical shear $\kappa_z^2 \ne 0$ and stable stratification in some direction $N^2_+ > 0$ there is some buoyantly unstable direction $N^2_- < 0$. The relation between \BVFs and the vertical shear helps to understand the linkage between GSF and COS modes, i.e.\ thermal instability and thermal overstability, and why their growth rates in the longer cooling time regime are proportional to each other.

One starts with the product of the two \BVFs $N^2_- N^2_+$. For that purpose we derive the extrema in $N^2$ using the normalized vector $\mathbf{k}$ with $k_R^2 + k_z^2 = 1$. We already derived the ratio $(k_R / k_z)_\pm$ in the main body of the paper, but for this proof it is easier to use an expression for $k_{R \pm}$. We start with the general expression for the directional \BVF given in Eq.\ \eqref{eq:BVTK} omitting the $\mathbf{k}$ for brevity:
\begin{eqnarray}
 N^2 = k_R^2 \left(N^2_z - N^2_R\right) + k_R \sqrt{1 - k_R^2} B + N^2_R.
  \label{eq:2dBV979},
\end{eqnarray}
which uses the short hand $B = \frac{c^2}{\gamma}(b_R s_z + b_z s_R))$.
We take the derivative with respect to $k_R$
\begin{eqnarray}
 \frac{\mathrm{d} N^2}{\mathrm{d} k_R} = 2 k_R \left(N^2_z - N^2_R\right) + \left(\sqrt{1 - k_R^2} - \frac{k_R^2}{\sqrt{1 - k_R^2}}\right) B = 0,
  \label{eq:2dBV984}
\end{eqnarray}
which leads to a quadratic equation in $k_R^2$:
\begin{eqnarray}
k_R^4  - k_R^2  + \frac{1}{4}\frac{B^2}{\left(N^2_z - N^2_R\right)^2 + B^2} = 0,
  \label{eq:2dBV984c}
\end{eqnarray}
with the two possible solutions:
\begin{eqnarray}
k_R^2 = \frac{1}{2} \pm \frac{1}{2}\sqrt{1 - \frac{B^2}{\left(N^2_z - N^2_R\right)^2 + B^2}}.
\end{eqnarray}
A check of the second derivative reveals the minimum to be at:
\begin{eqnarray}
k_{R-}^2 = \frac{1}{2} - \frac{1}{2}\sqrt{1 - \frac{B^2}{\left(N^2_z - N^2_R\right)^2 + B^2}}.
\label{EQ:D24}
\end{eqnarray}
A Taylor expansion for $B^2 \ll \left(N^2_z - N^2_R\right)^2$ leads to:
\begin{eqnarray}
k_{R-}^2 \approx  \frac{1}{4} \frac{B^2}{\left(N^2_z - N^2_R\right)^2 + B^2}.
\end{eqnarray}
We can furthermore approximate result by neglecting $B^2$ in the denominator and find:
\begin{eqnarray}
k_{R-} \approx  \frac{1}{2} \frac{B}{N^2_z - N^2_R},
\end{eqnarray}
which for small $k_{R-}/k_{z-}$ and $|N^2_z| > |N^2_R|$ is equivalent to the solution we found in Eq.\ \eqref{Eq:k_Rdk_za}.

But lets return to the full solution Eq.\ \eqref{EQ:D24}. It directly follows that:
\begin{eqnarray}
k_{z-}^2 = 1 - k_{R-}^2 = \frac{1}{2} + \frac{1}{2}\sqrt{1 - \frac{B^2}{\left(N^2_z - N^2_R\right)^2 + B^2}}.
\end{eqnarray}
The extremal \BVFs are now given via the orthogonal vectors $\mathbf{k}_- = (k_{R-}, k_{z-})$ and $\mathbf{k}_+ = (k_{z-}, -k_{R-})$.
Thus we can write the extremal \BVFs as:
\begin{eqnarray}
N^2_- = k_{R-}^2 \left(N^2_z - N^2_R\right)^2 + k_{R-} k_{z-} B + N_R^2,
\end{eqnarray}
and 
\begin{eqnarray}
N^2_+ = - k_{R-}^2 \left(N^2_z - N^2_R\right)^2 - k_{R-} k_{z-} B + N_z^2.
\end{eqnarray}
For readability we drop the $-$ in the following expression for $k$.
It will later be practical to determine the product of $k_R^2 k_z^2$, which is:
\begin{eqnarray}
k_R^2 k_z^2 =  \frac{1}{4}\left(\frac{B^2}{\left(N^2_z - N^2_R\right)^2 + B^2}\right).
\label{Eq:k_R2k_z2}
\end{eqnarray}
The product of both extremal \BVFs is then:
\begin{eqnarray}
N^2_+ N^2_-&=& -k_R^4 \left(N^2_z - N^2_R\right)^2 + k_R^2 k_z^2 B^2 + N_R^2 N_z^2\\
&+& k_R^2 \left(N^2_z - N^2_R\right)^2 - 2 k_R^2 \left(N^2_z - N^2_R\right) k_R k_z B + k_R k_z B \left(N^2_z - N^2_R\right).
\end{eqnarray}
As $k_R^4 = k_R^2 - k_R^2 k_z^2$ we find:
\begin{eqnarray}
N^2_+ N^2_-&=& -k_R^2 \left(N^2_z - N^2_R\right)^2 + k_R^2k_z^2 \left(N^2_z - N^2_R\right)^2  + k_R^2 k_z^2 B^2 + N_R^2 N_z^2\\
&+& k_R^2 \left(N^2_z - N^2_R\right)^2 - 2 k_R^2 \left(N^2_z - N^2_R\right) k_R k_z B + k_R k_z B \left(N^2_z - N^2_R\right).
\end{eqnarray}
Two terms involving $k_R^2$ cancel and we replace the first term $k_R^2k_z^2 () = - k_R^2k_z^2 ()+ 2 k_R^2k_z^2()$:
\begin{eqnarray}
N^2_+ N^2_-&=& - k_R^2 k_z^2 \left(N^2_z - N^2_R\right)^2 - k_R^2 k_z^2 B^2 + N_R^2 N_z^2\\
&+& 2 k_R^2 k_z^2 \left(N^2_z - N^2_R\right)^2 - 2 k_R^2 \left(N^2_z - N^2_R\right) k_R k_z B + k_R k_z B \left(N^2_z - N^2_R\right).
\end{eqnarray}
Inserting the expression for $k_R^2 k_z^2$ (Eq.\ \eqref{Eq:k_R2k_z2}) leads to
\begin{eqnarray}
N^2_+ N^2_-&=& - \frac{1}{4}B^2 + N_R^2 N_z^2 \label{Eq:B17}\\
&+& k_R k_z \left(N^2_z - N^2_R\right) \left[2 k_R k_z \left(N^2_z - N^2_R\right) + B (k_z^2 - k_R^2)\right].
\label{Eq:SBT1}
\end{eqnarray}
The first line will form the result that we are interested in, so we have to show that the second line is identical to zero, which is the case for 
$[...] = 2 k_R k_z \left(N^2_z - N^2_R\right) + B (k_z^2 - k_R^2) = 0$.
If we use the expressions for 
\begin{equation}
    k_z^2 - k_R^2 = \sqrt{1 - \frac{B^2}{\left(N^2_z - N^2_R\right)^2 + B^2}}.
\end{equation}
inside the square $[...]$ bracketed term of Eq.\ \eqref{Eq:SBT1} we find: 
\begin{equation}
    [...] = 2 k_R k_z \left(N^2_z - N^2_R\right) + B \sqrt{1 - \frac{B^2}{\left(N^2_z - N^2_R\right)^2 + B^2}},
\end{equation}
and by plugging in the expressions for $k_R k_z$ from Eq.\ (B13) one finds under the condition that $k_R k_z < 0$: 
\begin{equation}
[...] = - \sqrt{\frac{\left(N^2_z - N^2_R\right)^2}{\left(N^2_z - N^2_R\right)^2 + B^2}} + \sqrt{\frac{\left(N^2_z - N^2_R\right)^2 + B^2}{\left(N^2_z - N^2_R\right)^2 + B^2} - \frac{B^2}{\left(N^2_z - N^2_R\right)^2 + B^2}} = 0.
\end{equation}
We find that $[...] = 0$ and thus can be dropped from Eq.\ \eqref{Eq:B17} and we get:
\begin{eqnarray}
N^2_+ N^2_-&=& - \frac{1}{4}B^2 + N_R^2 N_z^2.
\end{eqnarray}
Using the expressions for $B^2$, $N_R^2$ and $N_z^2$, we find
\begin{eqnarray}
N^2_+ N^2_-&=& - \frac{c^2}{4 \gamma}\left[\left(b_R s_z + b_z s_R\right)^2 - 4 b_R s_R b_z s_z\right] \\
&=& - \frac{c^2}{4 \gamma}\left(b_R s_z - b_z s_R\right)^2\\
&=& - \frac{1}{4}\kappa_z^4 \,\,\,\,\,\,\,\,\,\mathbf{QED.}
\end{eqnarray}
More generally one can directly formulate:
\begin{eqnarray}
N^2_+ N^2_-&=& - \kappa_-^2 \Omega^2.
\end{eqnarray}
For the case of a vertically isothermal stratification and sufficient distance to the midplane, the most stable \BVF is in the vertical direction $N_+^2 \approx N_z^2$, which then leads to the approximate result in Section \ref{SS:AnalyticExpressions}:
\begin{eqnarray}
N^2_-&=&  - \frac{\kappa_z^4}{4 N_z^2}.
\end{eqnarray}
For the particular height at which the most stable \BVF is exactly vertical $N_+^2 = N_z^2$, we then find that $N_-^2 = N_R^2$.


\end{document}